\documentclass[journal]{IEEEtran}
\usepackage{stfloats}
\usepackage{amsmath,amssymb,amsfonts}
\usepackage{algorithmic}
\usepackage{graphicx}
\usepackage{textcomp}
\def\BibTeX{{\rm B\kern-.05em{\sc i\kern-.025em b}\kern-.08em
		T\kern-.1667em\lower.7ex\hbox{E}\kern-.125emX}}

\usepackage{color,soul}
\usepackage{multirow,slashbox,booktabs}
\usepackage{tabularx}
\usepackage{bm}

\begin{document}
    % \title{Enhancing PIR-based Multi-person Localization through Combining Deep Learning with Domain Knowledge}
    \title{A Deep-learning-based Method for PIR-based Multi-person Localization}
    
    \author{Tianye Yang, Peng Guo, Wenyu Liu, Xuefeng Liu, Tianyu Hao}

%    \affiliation{%
%    	\institution{ School of Electronic and Infor. Engi., Huazhong Univ. of Sci. and Tech.}}
%    \email{d201677515@hust.edu.cn
%    }
    
%    \author{Peng Guo}
%    \affiliation{%
%    	\institution{School of Electronic and Information Engineering, Huazhong University of Science and Technology}}
%    \email{guopeng@hust.edu.cn}	
%    
%    \author{Wenyu Liu}
%    \affiliation{%
%    	\institution{School of Electronic and Information Engineering, Huazhong University of Science and Technology}}
%    \email{liuwy@hust.edu.cn}	
    
%    \author{Xuefeng Liu}
%    \affiliation{%
%    	\institution{State Key Lab. of VR Tech. \& Sys.}
%    	\institution{School of C.S.E., BeiHang University}}
%    \email{liu_xuefeng@buaa.edu.cn}	
    
%    \author{Tianyu Hao}
%    \affiliation{%
%    	\institution{School of Electronic and Information Engineering, Huazhong University of Science and Technology}}
%    \email{hty@hust.edu.cn}	
    
  \maketitle
	
	\begin{abstract}
		Device-free localization (DFL) based on pyroelectric infrared (PIR) sensors has attracted much attention due to its advantages of low cost, low power consumption, and privacy protection. However, most existing PIR-based DFL methods require high deployment density to achieve high localization accuracy. Recently, a few works proposed that the deployment density can be reduced through deeply analyzing the analog output of PIR sensors. However, these methods can not well handle the localization task in multi-person scenarios yet. 
		In this paper, we propose a novel neural network for PIR-based multi-person localization, which appropriately leverages a series of domain knowledge. Specifically, the proposed network consists of two modules: one is for determining the number of persons and another is for determining their locations. Meanwhile, the module of person counting is further designed as a two-stage network: one stage is for signal separation and another is for single-person detection. The module for localization is also designed as a two-stage network: one stage is for signal extraction and another is for single-person localization. 
		Through the above methods, we succeed to remarkably reduce the deployment density of the traditional PIR-based method by about 76\%, while maintaining the localization accuracy.
	\end{abstract}
	
%	\keywords{Device-free localization (DFL); PIR sensors; deep learning; domain knowledge.}
	\IEEEpeerreviewmaketitle
	
%	\titlepgskip=-15pt

	\section{Introduction}
	\label{sec:Introduction}
	
	\IEEEPARstart{D}{evice-free} localization (DFL) has attracted attention from a lot of researchers, since it is required in many practical applications, such as intruder tracking and health caring \cite{shit2019ubiquitous}. There has been various DFL systems which are respectively based on cameras \cite{yang2007robust}, RF devices \cite{xie2019md}, light sensors \cite{mao2011ilight}, acoustic sensors \cite{guo2011localising}, electric field sensors \cite{grosse2016platypus} or pyroelectric infrared (PIR) sensors. Among the above systems, the PIR-based system is promising due to its advantages of low cost, low power consumption, and privacy protection. Typically, the price of a PIR sensor is less than 1 dollar, and the power consumption of a PIR sensor is only about 50$\mu W$ \cite{ManualOfPIR}.
	
	However, existing PIR-based systems still have a defect of high deployment density. The main reason of the high deployment density is that most existing systems utilize PIR sensors as binary motion detectors, and discard abundant location information contained in the PIR sensors' analog output. Recently, some works \cite{monaci2012indoor, narayana2015pir, yang2019push} report that the deployment density of PIR-based localization system can be reduced through modelling the relationship between PIR sensors' raw output and a person's location. Although these methods are promising, they can not well handle the multi-person scenarios yet. A main reason is that, the relationship between multi-person locations and the raw output is complex and difficult to be manually modelled. Therefore, we intend to utilize deep neural networks to solve the PIR-based multi-person localization task, since it has been demonstrated as a powerful tool that can automatically model complex relationships in tasks such as image classification \cite{krizhevsky2012imagenet,simonyan2014very,he2016deep} and speech recognition \cite{hinton2012deep,graves2013speech,amodei2016deep},

	A straightforward deep learning solution for the PIR-based multi-person localization is to input the raw signals of PIR sensors into a neural network and let the network output the positions of target persons and the corresponding possibilities. The network architecture could be chosen as sequential, such as BiLSTM \cite{graves2013hybrid}, which is good at modeling sequential data. Although the above solution seems decent, we demonstrate that its performance is not satisfactory. 
	To improve the performance, we try to utilize the domain knowledge to design a task-specific network architecture and a series of techniques of signal preprocessing and data augmentation.

\begin{figure}[h]
			\centering
			\includegraphics[width=0.65\linewidth]{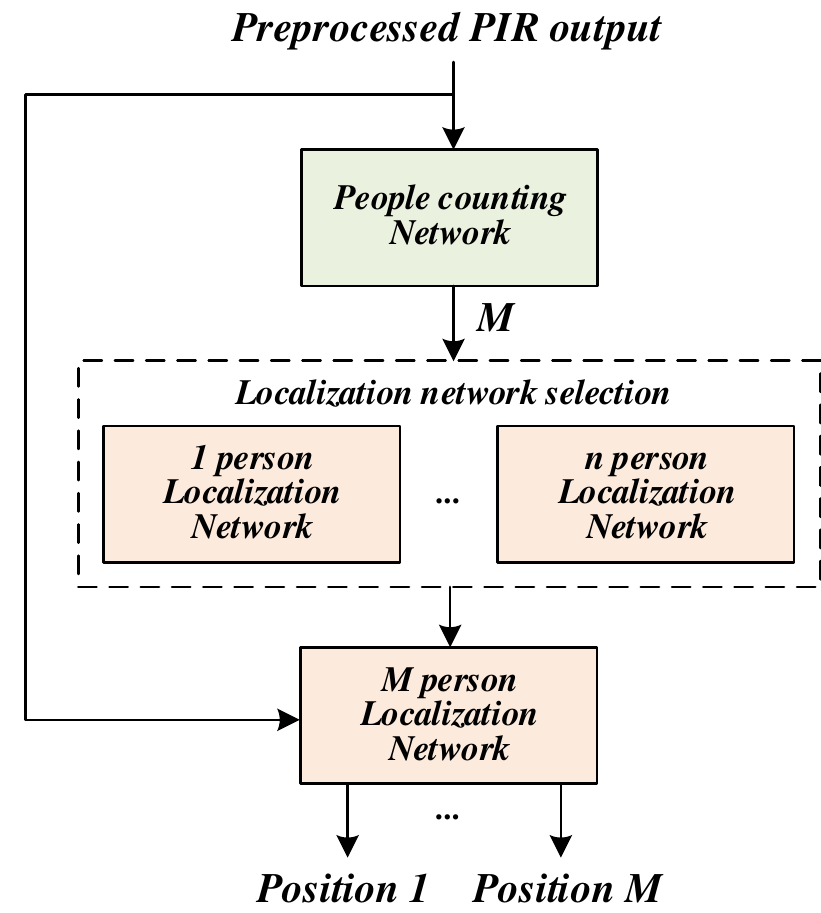}
			\caption{PIRNet: a two-stage architecture for multi-person localization.}%
			\label{fig:Framework of PIRNet}
			%\vspace{-0.4em}
		\end{figure}

	First, we propose an architecture PIRNet specially designed for the PIR-based multi-person localization. This design utilizes the ideas of modular learning \cite{lu1999task} and multi-stage learning \cite{gulccehre2016knowledge}. Specifically, modular learning proposes not to assign a complex task to a single network, but to decompose the complex task into several sub-tasks and utilize several individual networks to solve them. The rationality of this approach is that the decomposition could introduce domain knowledge and reduce the search space of the training procedure \cite{rempis2010search}. Based on this idea, we design the architecture of PIRNet as shown in Fig. \ref{fig:Framework of PIRNet}. It can be seen that we divide the multi-person localization task into two sub-tasks of person counting and location prediction, and utilize two networks to solve them respectively. Besides, to further improve the performance of the networks for person counting and localization, we design their structures utilizing the idea of multi-stage learning. Similar to modular learning, multi-stage learning proposes to decompose a complex task into sub-tasks and utilize a hierarchical network to solve them. Meanwhile, the hierarchy contains multiple tiers, which are utilized to solve different sub-tasks. Different from modular learning, the input of a tier of the multi-stage network is the output of its previous tier. Based on this idea, we design the person counting network as a two-stage network, one stage for signal separation and the other for single-person detection. As for the network for location prediction, we also design it as two-stage, one stage for signal extraction and the other for single-person localization. The above two-stage design is motivated by the domain knowledge of that the signal of multi-person approximately equals to the addition of signals of each single person and some noise. More details are introduced in Section \ref{sec:Principle of PIR sensor} and Section \ref{sec:Detailed structure of PIRNet}.
	
%	\Figure[t!](topskip=0pt, botskip=0pt, midskip=0pt)[width=2 in]{Fig25_PIRNet_architecture.pdf}
%	{\bf PIRNet: a two-stage architecture for multi-person localization.\label{fig:Framework of PIRNet} \vspace{-3.5em}}
	
	Second, instead of directly utilizing the raw output signals of PIR sensors as the network's input, we propose to utilize the preprocessed signals as the input. There are two reasons for the preprocessing. First, the raw signal usually suffers from ambient noise which could decrease the network's performance. Second, the change of environmental temperature influences the amplitude of the raw signal and could further decrease the robustness of the trained network. To alleviate the negative influences of the noise and environmental temperature change, we propose a scheme of denoising and a scheme of normalization, which are introduced in Section \ref{sec:preprocessing} in detail.

	Third, we propose two data augmentation method to achieve higher person counting and localization performance. Specifically, based on the physical model of PIR sensors, we at first propose a method to simulate training samples with lower or higher speed by stretching or compressing the initial samples. Then, we propose another method to simulate training samples influenced by surrounding heat sources through randomly amplifying or suppressing the initial samples. After adding these simulated training samples into the initial dataset, the performance of person counting and localization can also be improved. The details will be introduced in Section \ref{sec:data augmentation}.

	The contributions of this paper are as follows:
	\begin{enumerate}
		\item We first build a signal model of a PIR sensor when multiple persons move simultaneously, and find that the signal of multi-person is separable. Base on this separability, we propose a task-specified network architecture PIRNet for PIR-based multi-person localization. With this method, we achieve average localization errors of 0.43m, 0.62m, and 0.82m in scenarios where 1, 2, and 3 persons simultaneously move.

		\item We analyze the influence of a series of practical factors to the signal of a PIR sensor, and then propose the corresponding solutions. Specifically, we propose a data preprocessing strategy and two data augmentation strategies to improve the person counting and localization accuracy of PIRNet. With these schemes, the person counting accuracy is increased from 94.2\% to 96.1\%, and the average localization error is decreased from 0.72m to 0.62m. The detailed results are introduced in Section \ref{sec:Improvement through preprocessing and data augmentation}.
		
		\item Compared with the traditional PIR-based methods, we reduce the deployment density from $0.34\ sensor/m^2$ to $0.08\ sensor/m^2$, while maintaining the localization accuracy. The detailed comparison is introduced in Section \ref{sec:Comparison with other PIR-based methods}.
	\end{enumerate}
	
	The rest of this article is organized as follows. In Section \ref{sec:Related works}, we summarize the related works about PIR-based localization. In Section \ref{sec:Principle of PIR sensor}, we briefly introduce the working principle of PIR sensors. In Section \ref{sec:Detailed structure of PIRNet}, the detailed structures of the PIRNet and the baseline networks for comparison are introduced. In Section \ref{sec:Preprocessing and data augmentation}, we introduce the methods of preprocessing and data augmentation. In addition, the experiments for validating the proposed approach are introduced in Section \ref{sec:evaluations}, followed by conclusions and some discussions in Section \ref{sec:conclusion}.

	\section{Related works}
	\label{sec:Related works}
	
\begin{figure}[h]
		\vspace{-0.5em}
		\centering
		\includegraphics[width=0.45\linewidth]{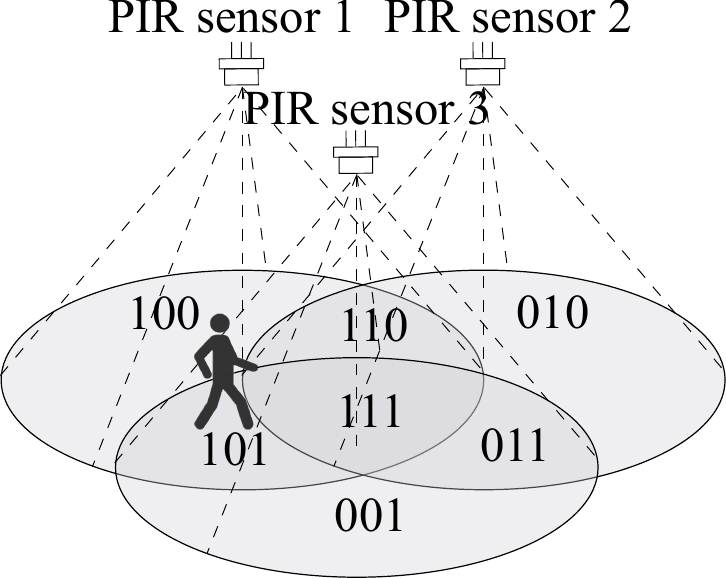}
		\vspace{-0.3em}
		\caption{\bf Basic idea of traditional PIR-based localization systems.}%
		\label{fig:Basic idea of binary systems}
		\vspace{-0.5em}
	\end{figure}

	%\subsection{PIR-based localization}
	Fig.\ref{fig:Basic idea of binary systems} illustrates the basic idea of most traditional PIR-based localization methods. First, one needs to deploy a lot of PIR sensors in an environment and let their detection zones partially overlapped. Then, when a person moves in different zones, different groups of PIR sensors will be triggered. 
	For example, when a person moves in the zone `101', PIR sensors 1 and 3 will be triggered and report `1'. PIR sensors 2 will not be triggered and report `0'. Therefore, through the binary numbers reported by each PIR sensors, we can determine the zone where the person is. 
	
	The early work about PIR-based localization is done by Hao et al. in \cite{hao2006human}. In this work, the authors propose a preliminary method for localizing a single-person. In \cite{hao2009multiple}, Hao et al. extends the above method to multi-person scenarios through a dense deployment strategy. Specifically, they first build a probability model of the multi-person locations and the PIR sensors' alarm sequences. Then, they utilize the EM algorithm to get a maximum posterior probability estimation of the locations. In \cite{lu2016preprocessing}, Lu et al. further improve the localization accuracy of the multi-person scenario through schemes of sensor selection and calibration. In \cite{yang2014novel}, Yang et al. propose a deployment scheme that all PIR sensors are deployed on the ground by equipping PIR sensors with cone optics. Subsequently, in \cite{yang2015novel}, they extend the above method to the multi-person scenario through a coarse-fine method. To further improve the localization accuracy, Yang et al. propose another two enhanced methods of inconstant confidence \cite{yang2016credit} and virtual detection lines \cite{yang2017multiple}. On the other hand, Luo et al. \cite{luo2017simultaneous} propose a method that allows deploying PIR sensors on the ceiling. The above methods mainly utilize a PIR sensor as a binary motion detector. These methods have a common defect of high deployment density. The reason is that their localization accuracy is mainly dependent on the granularity of the overlapped zones. 
	
	To achieve high localization accuracy with lower deployment density, researchers propose some new methods based on deeply analyzing the raw output of PIR sensors. In \cite{monaci2012indoor}, Monaci et al. propose a method to estimate the direction of a person, and then utilized the estimated direction to achieve single-person localization. The defect of this method is that it is tested in a narrow range of about $6m \times 3m$ and seems not applicable in a larger range. The reason is that the method utilizes the amplitude of a PIR sensor's output to distinguish the direction. However, in practice, the amplitude of the output is not only determined by the absolute direction, but also the distance between the person and the sensor. Therefore, in a larger range, the distance may vary in wide range, making the relationship between the amplitude and the direction of a person not constant.
	
	In \cite{zappi2010tracking}, Zappi et al. propose that the distance of a person to a PIR sensor can be coarsely estimated by the amplitude of its raw out. Subsequently, Narayana et al. \cite{narayana2015pir} propose a method to achieve a more accurate estimation of distance, and then achieve single-person localization through the estimated distance. The limitation of this method is that the required deployment density is still high and the required PIR sensors are customized. The reason of this limitation still lies that the amplitude depends both on the distance and direction of a person to the sensor. To decrease the ambiguity caused by the direction change, the method requires 8 customized PIR sensors which have different detection ranges to cover a range of about 50$m^2$.
	
	In \cite{yang2019push}, instead of utilizing the amplitude of a PIR sensor's output, Yang et al. propose a method that first estimates the azimuth change of a person through the short-time frequency of the output, and then achieves more accurate single-person localization with fewer PIR sensors. However, although this method achieves good performance in the single-person scenario, it can not well handle the multi-person scenario yet. 
	
	\section{Preliminary knowledge of PIR sensors}
	\label{sec:Principle of PIR sensor}
	
	The PIR sensor is commonly used for person detection \cite{odon2010modelling}. In practice, it is usually combined with a Fresnel lens array to enlarge its detection range and sensitivity \cite{cirino2006design}. For simplicity, the `PIR sensor' in this article refers to `a PIR sensor combined with a Fresnel lens array'.

\begin{figure}[h]
		\centering
		\includegraphics[width=0.6\linewidth]{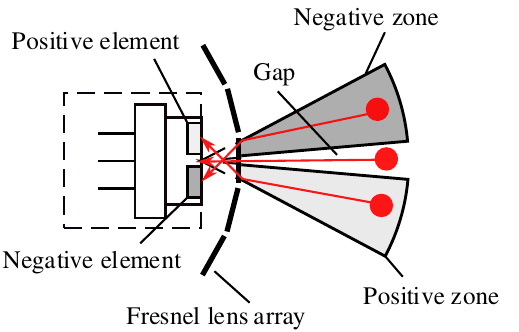}
		\caption{\bf The principle of a PIR sensor.}%
		\label{fig:pir_sensor_principle}
		\vspace{-0.5em}
	\end{figure}

\begin{figure}[h]
		\centering
		\includegraphics[width=0.7\linewidth]{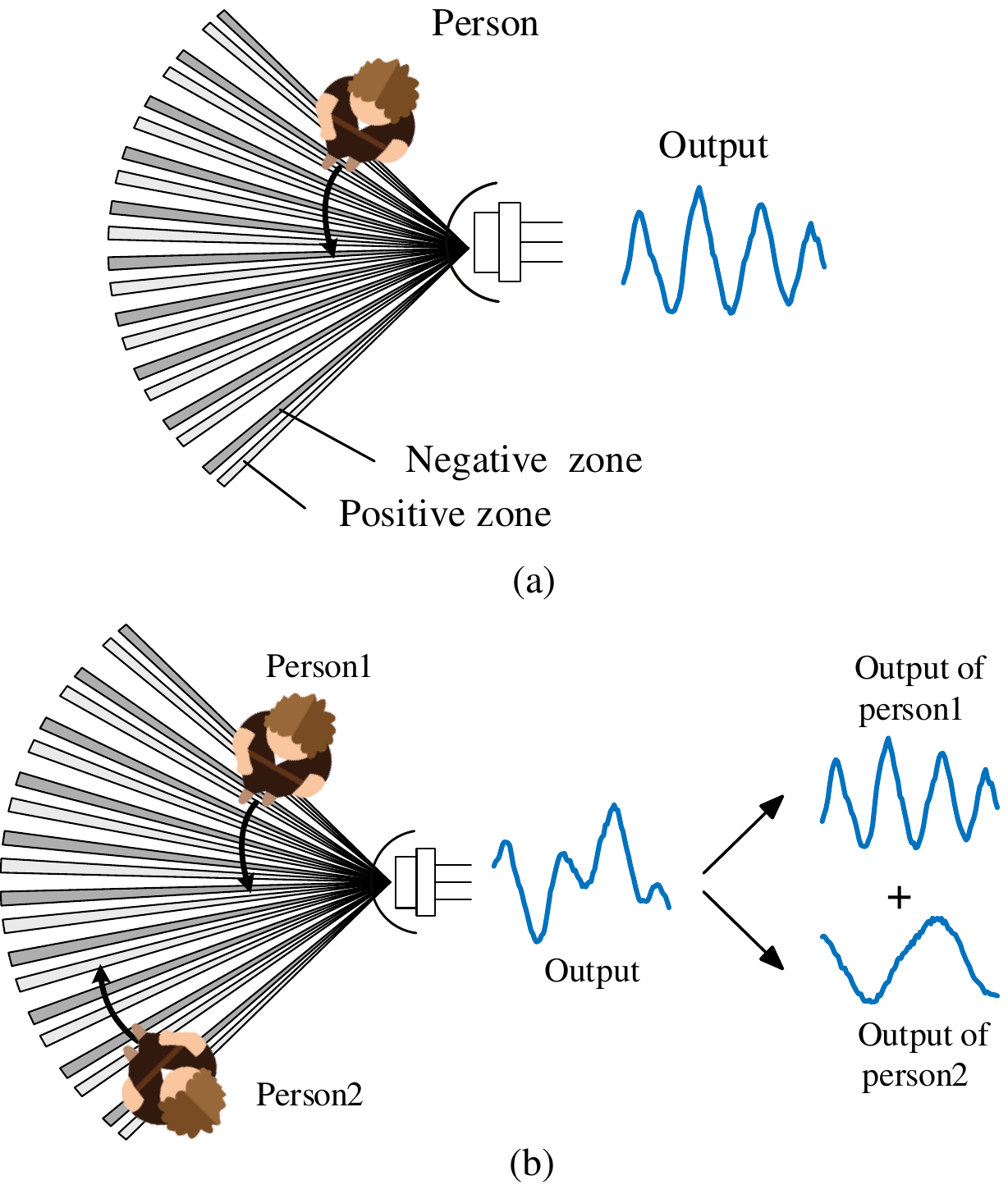}
		\caption{\bf The DHF of a PIR sensor: (a) when one person moves in front of it; (b) when two person move in front of it.}%
		\label{fig:single-person-signal}
		\vspace{-1.5em}
	\end{figure}

	A PIR sensor senses a person through the infrared radiation emitted by him/her. The principle of a PIR sensor is illustrated in Fig. \ref{fig:pir_sensor_principle}. A PIR sensor contains a positive element and a negative element, and the sensor's output reflects the differential between the heat fluxes (DHF) on its positive and negative elements. The DHF is influenced by the positions of a person. Specifically, a PIR sensor's detection zone can be divided into many fan-shaped zones of two kinds, the positive zones and the negative zones. When a person locates in the positive zone, his/her infrared radiation is concentrated on the sensor's positive element, and the heat flux on the positive element rises. In contrast, when a person locates in the negative zone, his/her infrared radiation is concentrated on the negative element, and the heat flux on the negative element rises. Besides, when a person locates in the gaps between the positive and negative zones, the radiation is concentrated neither on the positive element nor the negative element, and would not influence the heat fluxes on the positive/negative elements. Therefore, when a person moves in front of a PIR sensor, the DHF of the PIR sensor will increase and decrease alternatively like a sine wave, as illustrated in Fig. \ref{fig:single-person-signal}(a).
	
	Furthermore, when multiple persons move simultaneously, the sensor's DHF approximately equals to the summarization of the their DHFs when they move individually, as illustrated in Fig. \ref{fig:single-person-signal}(b). This phenomenon is supported by the law of photometry which is introduced in \cite{budzier2011thermal}. Specifically, the authors introduce that the heat flux on a surface $A_1$, which receives the radiation emitted by a surface $A_2$, can be calculated as follows:
	\begin{equation}
	{\phi _1} = \int {\frac{{\sigma {\omega _{21}}}}{\pi }} ({T_1}^4 - {T_2}^4)d{A_2}
	\label{eqn:law of photometry}
	\end{equation}
	where $\sigma$ is the Stefan--Boltzmann constant; $T_1$ and $T_2$ are the temperatures of the surface $A_1$ and $A_2$; $\omega_{21}$ is the projected solid angle of area $A_2$ in relation to area $A_1$. Therefore, assuming that the persons do not occlude each other, the DHF of a PIR sensor can be calculated as follows:
	\begin{equation}
	{\phi _d} = \sum\limits_{i = 1}^N {{\phi _d}^{(i)}} + {\phi _n}
	\label{eqn:DHF_no_blocking}
	\end{equation}
	where $N$ is the total number of persons; ${\phi _n}$ is a noise term that represents the DHF caused by surrounding heat sources; ${\phi_d}^{(i)}$ is the DHF caused by the $i$th person, which is calculated as follows:
	\begin{equation}
	{\phi _d}^{(i)} = \int {\frac{{\sigma {\omega _{iP}}}}{\pi }} ({T_P}^4 - {T_i}^4)d{A_i} - \int {\frac{{\sigma {\omega _{iN}}}}{\pi }} ({T_N}^4 - {T_i}^4)d{A_i}
	\label{eqn:DHF_single_person}
	\end{equation}
	where $A_i$ represents the surface of the $i$th person; $\omega_{iP}$ represents the projected solid angle of $A_i$ in relation to the surface of the sensor's positive element; $\omega_{iN}$ represents the projected solid angle of $A_i$ in relation to the surface of the sensor's negative element; $T_P$ and $T_N$ are the surface temperatures of the positive and negative elements; $T_i$ is the surface temperature of the $i$th person.
	
	In practice, when multiple persons move simultaneously, they would occlude each other, and some of their radiation could not achieve the positive/negative elements. Therefore, we should add another term into (\ref{eqn:DHF_no_blocking}) to represent the distortion caused by the occlusion effect:
	\begin{equation}
	{\phi _d} = \sum\limits_{i = 1}^N {{\phi _d}^{(i)}}  + {\phi _o} + {\phi _n}
	\label{eqn:DHF_blocking}
	\end{equation}
	where $\phi_{o}$ represents the distortion caused by the occlusion effect. Since we assume that the trajectories of the persons are random and independent, $\phi_{o}$ could also be considered as a noise term.

	In addition, we should notice that the output of a PIR sensor is not the DHF in practice. Specifically, a PIR sensor can be considered as a second-order dynamic system whose input is the DHF. Its transfer function is as follows \cite{yang2019push}:
	\begin{equation}
	V(s) = \frac{{As}}{{B{s^2} + Cs + 1}} \cdot {\phi _d}(s)
	\label{eqn:transform function of dual-element PIR sensor}
	\end{equation}
	where $V(s)$ is the Laplace transform of the PIR sensor's output voltage; ${\phi _d}(s)$ is the Laplace transform of the sensor's DHF; $A,B,C$ are some constants related to the physical characters of the sensor.

	\section{The Proposed PIRNet Scheme}
	\label{sec:Detailed structure of PIRNet}
	In this section, we introduce the proposed PIRNet in details, which leverages some domain knowledge to the neural network design. 
	As introduced in Section \ref{sec:Introduction}, PIRNet contains two networks respectively for person counting and localization. To improve the performance of these two networks, we design them as two-stage networks. 
	
	\subsection{Network for person counting}
	\label{sec:Network for person counting}
\begin{figure}[h]
		\centering
		\includegraphics[width=0.85\linewidth]{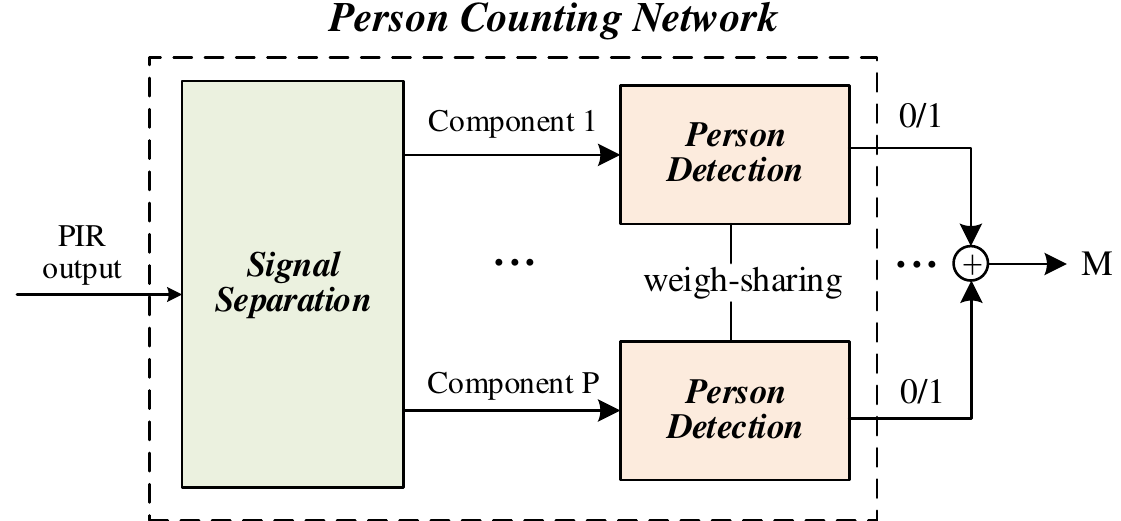}
		\caption{\bf Framework of the network for person counting.}%
		\label{fig:Framework of people counting network}
		\vspace{-0.5em}
	\end{figure}

	As illustrated in Fig. \ref{fig:Framework of people counting network}, the network for person counting contains two stages. The first stage is utilized to separate the network's input into $P$ individual components, each of which is corresponding to a person or a noise source. The second stage is utilized to determine whether a separated component is corresponding to a person or a noise source. Finally, the number of components corresponding to persons is considered as the prediction of the number of persons. The motivation behind this design is that the output of the multi-person scenario approximately equals the outputs of every single person. Therefore, the task of person counting can be divided into two sub-tasks of signal separation and binary classification of whether a separated component is corresponding to a person. It should be noted that the output of the stage for signal separation is not supervised. Therefore, the separated components are not the components of the PIR sensors' real signal, but their abstract expression.
	
\begin{figure}[h]
		\centering
		\includegraphics[width=0.5\linewidth]{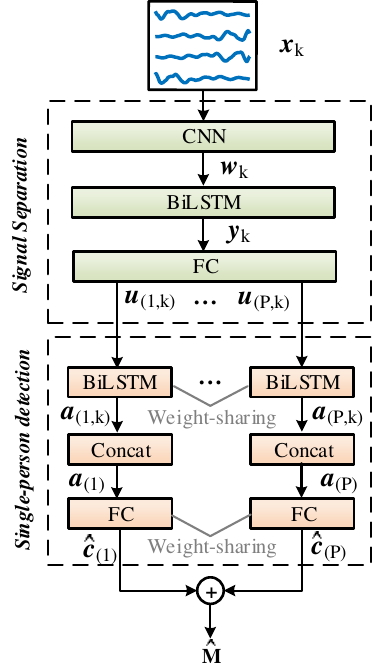}
		\caption{\bf Detailed structure of the network for person counting.}%
		\label{fig:network for people counting}
		\vspace{-1.5em}
	\end{figure}

	The detailed structure of the person counting network is illustrated in Fig. \ref{fig:network for people counting}. 
	
	\subsubsection{Input}
	The input of this network is the preprocessed raw output of $N$ PIR sensors (the procedure of preprocessing will be introduced in Section \ref{sec:preprocessing}). The duration of the input signal is $D$. Since we utilize recurrent layers in the network, we divide the input into $K$ segments. In the following, we use $\bm{x}_{k} \in {\mathbb{R}^{N \times T}}$ to represent the $k$th segment. Meanwhile, the $n$th row of $\bm{x}_{k}$ represents the DHF of the $n$th PIR sensor, and $T$ is the length of each divided segment.
	
	\subsubsection{Separation}
	The structure of the separation stage borrows the idea of Tasnet \cite{luo2018tasnet}, which is a popular architecture for speech separation. Specifically, it contains a 1-D gated convocation layer \cite{dauphin2017language}, two BiLSTM layers \cite{graves2013hybrid} and one fully connected layer. 
	
	The output of the 1-D gated convolution layer is calculated as follows:
	\begin{equation}
	{\bm{w}_k} = ReLU(\bm{x}_k * \bm{u}) \odot Sigmoid(\bm{x}_k * \bm{v}) + \bm{b}
	\end{equation}
	where $\bm{u} \in {\mathbb{R}^{T \times C}}$, $\bm{v} \in {\mathbb{R}^{T \times C}}$ and $\bm{b} \in {\mathbb{R}^{N \times C}}$ are learnable parameters, $C$ is the number of convolution kernels; $*$ represents the operation of matrix multiplication; $\odot$ represents the operation of element-wise matrix multiplication; the $ReLU(\cdot)$ and $Sigmoid(\cdot)$ denote the ReLU and Sigmoid activation function \cite{lecun2015deep}. 
	
	The calculation of the BiLSTM layers is as follows. We utilize $\bm{y}_{(i,k)}$ to represent the output vector of the $i$th BiLSTM layer, and it is calculated as follows:
	\begin{equation}
	{{\bm{y}}_{(i,k)}} = \left\{ {\begin{array}{*{20}{l}}
		{flatten({{\bm{w}}_k})*{\bm{I}_1},i = 0;}\\
		{H({{\bm{y}}_{(i - 1,k)}},{{\bm{s_{f1}}}_{(i,k - 1)}},{{\bm{\theta }}_{f1}}^{(i)}) + }\\
		{H({{\bm{y}}_{(i - 1,K - k+1)}},{{\bm{s_{b1}}}_{(i,K - k)}},{{\bm{\theta }}_{b1}}^{(i)}),i > 0.}
		\end{array}} \right.
	\end{equation}
	where $\bm{y}_{(0,k)}$ represents the input of the first layer; $flatten(\cdot)$ represents the flattening operation; $\bm{I}_1$ is a learnable input matrix; ${{\bm{\theta}_{f1}} }^{(i)}$ and ${\bm{\theta_{b1}} }^{(i)}$ are the learnable parameters of the forward and backward cells of the $i$th layer; $H(\cdot)$ represents the function to calculate the layer output, whose detailed definition can be found in \cite{LSTM}; ${{{\bm{s}_{f1}}}_{(i,k)}}$ and ${{{\bm{s}_{b1}}}_{(i,k)}}$ respectively represent the historical states of the forward and backward LSTM cells of the $i$th layer at the $k$th time step. In addition, the activation functions of the LSTM cells are set to ReLU. The initial states of all cells are set to zero. To improve the convergence speed, we also perform layer normalization on the LSTM cells \cite{ba2016layer}.
	
	Assuming the output of the last BiLSTM layer is ${\bm{y}}_{k}$, the output of the fully connected layer is as follows:
	\begin{equation}
	[{{\bm{u}}_{(1,k)}},...,{{\bm{u}}_{(P,k)}}] = ReLU({{\bm{y}}_{k}} * {{\bm{W}}_{fc1}}) + {{\bm{b}}_{fc1}}
	\end{equation}
	where $\bm{u}_{(p,k)}$ represents the $p$th separated component; ${{\bm{W}}_{fc}}$ and ${{\bm{b}}_{fc}}$ are respectively the weights and bias of the fully connected layer, which are also learnable.

	\subsubsection{Single-person Detection}
	The stage for single-person detection receives a separated component of the first stage as the input and then outputs the possibility that this component is related to a person. Specifically, it contains 2 BiLSTM layers, a concatenation layer, and a fully connected layer. 
	
	The output of the BiLSTM layers is calculated as follows:
	\begin{equation}
	{{\bm{a}}_{(i,p,k)}} = \left\{ {\begin{array}{*{20}{l}}
		{{{\bm{u}}_{(p,k)}}*{{\bm{I}}_2},i = 0;}\\
		{H({{\bm{a}}_{(i - 1,p,k)}},{{\bm{s}}_{{f}2}}_{(i,p,k - 1)},{{\bm{\theta }}_{{{f}}2}}^{(i)}) + }\\
		{H({{\bm{a}}_{(i - 1,p,K-k+1)}},{{\bm{s}}_{{{b}}2}}_{(i,p,K - k)},{{\bm{\theta }}_{{{b}}2}}^{(i)}),i > 0.}
		\end{array}} \right.
	\end{equation}
	where ${{\bm{a}}_{(0,p,k)}}$ is the input; $\bm{I}_2$ is the input matrix, which is learnable; ${{\bm{a}}_{(i,p,k)}}$ is the output of the $i$th BiLSTM layer;  ${{\bm{\theta}_{f2}} }^{(i)}$ and ${{\bm{\theta}_{b2}} }^{(i)}$ are the learnable parameters of the forward and backward cells of the $i$th layer; ${{{\bm{s}_{f1}}}_{(i,k)}}$ and ${{\bm{s_{b1}}}_{(i,k)}}$ are respectively the historical states of the forward and backward LSTM cells. The initial states of all cells are also set to zero.
	
	The concatenation layer concatenates ${{\bm{a}}_{(2,p,k)}}$ $(k=1,...,K)$ into a vector ${\bm{a}}_{(p)}$. Then, ${\bm{a}}_{(p)}$ will be sent into the fully connected layer. The width and activation function of the fully connected layer is respectively 1 and Sigmoid. The output of the fully connected layer is referred as ${\hat c}_{(p)}$, which is the probability that the separated component ${\bm u}_{(p,k)}$ $(k=1,...,K)$ belongs to a moving person. 
	
	\subsubsection{Loss function}
	The loss function of the network is as follows:
	\begin{equation}
	l = \sum\limits_{p = 1}^P {\sqrt {{{({{\hat c}_{(p)}} - {c_{(p)}})}^2}} } 
	\label{eq:loss_function_count}
	\end{equation}
	where ${{c}}_{(p)}$ is the true possibility that the $p$th separated component is related to a person.
	
	In addition, when determine ${{c}}_{(p)}$, we need to consider its character of permutation invariant. For example, assuming $P=3$ and the number of moving persons is 2, the correct label of the person counting network can be any of `${\hat c}_{(1)}=1, {\hat c}_{(2)}=1, {\hat c}_{(3)}=0$', `${\hat c}_{(1)}=1, {\hat c}_{(2)}=0, {\hat c}_{(3)}=1$', and `${\hat c}_{(1)}=0, {\hat c}_{(2)}=1, {\hat c}_{(3)}=1$'. Therefore, we adopt permutation invariant training (PIT) \cite{yu2017permutation}. The procedure of PIT contains two steps. For a specific input sample, we first calculate the corresponding losses of all possible labels. Then, we choose the label of the lowest loss as the label for error back propagation.
	
	\subsubsection{Hyperparameters}
	
	In our experiments, the period of input signal $D$ is set to 2.5s. The sampling frequency of each PIR sensor is 60Hz. The maximum person number $P$ is set to 3, which is the same as the related state-of-the-art methods. The segments number $K$ of each input is set to 5. The number of 1d-convolution kernel, i.e. $C$, is set to 8. The width of the BiLSTM layers in stage for separation is set to 64. The width of the fully connected layer in the stage for separation is set to 12. The width of the BiLSTM layers in stage for single-person detection is set to 16.
	
	\subsection{Network for localization}
\begin{figure}[h]
		\centering
		\vspace{-0.5em}
		\includegraphics[width=0.85\linewidth]{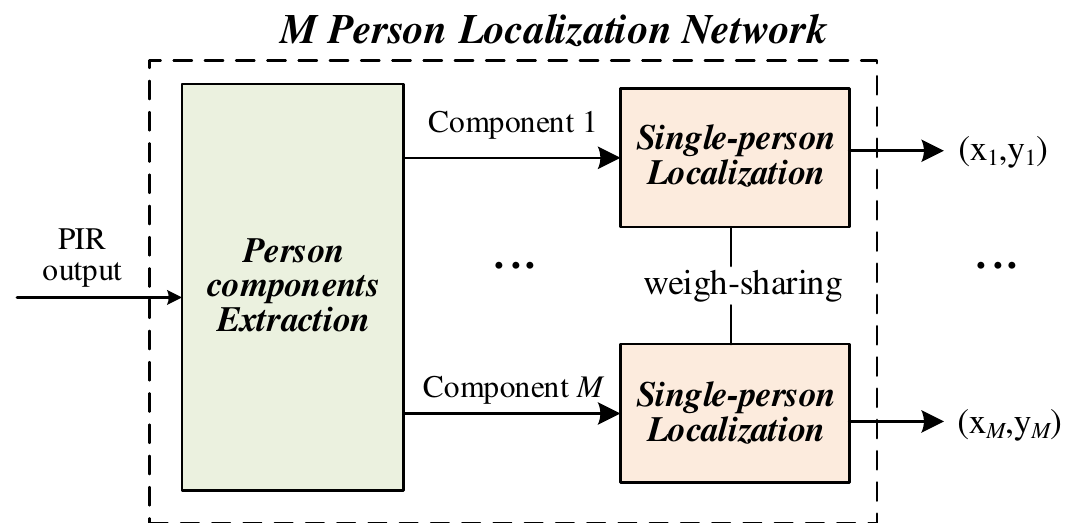}
		\caption{\bf Framework of the network for localizing multi-person of fixed number M.}%
		\label{fig:network for localizing multi-person}
		\vspace{-0.5em}
	\end{figure}

	As illustrated in Fig. \ref{fig:network for localizing multi-person}, the localization network also contains two stages. The first stage is utilized to extract the components of $M$ persons contained in the PIR sensors' output. The second stage is utilized to predict the location of each person through the corresponding extracted components. In addition, the output of the stage for signal extraction is also not supervised.
	
\begin{figure}[h]
		\centering
		\includegraphics[width=0.5\linewidth]{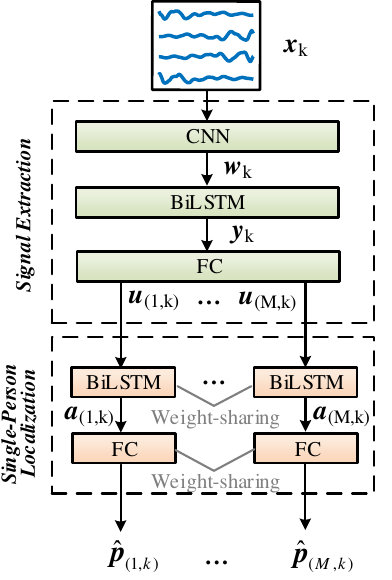}
		\caption{\bf Detailed structure of the network utilized for localizing multi-person of fixed number.}%
		\label{fig:network for localization}
		\vspace{-1.5em}
	\end{figure}

	Fig. \ref{fig:network for localization} illustrates the detailed structure of the network for localization. The first stage contains a 1-D gated convocation layer \cite{dauphin2017language}, two BiLSTM layers \cite{graves2013hybrid} and a fully connected layers. The input of this network is also $\{ \textbf{x}_{1}, ..., \textbf{x}_{K} \}$. In addition, the second stage contains two BiLSTM layers and 1 fully connected layer. 
	The calculation this network is similar to the above person counting network and we no longer repeat them here.
	
	\subsubsection{Loss function}
	
	The loss function of this network is as follows:
	\begin{equation}
	l = \sum\limits_{m = 1}^M {\sum\limits_{k = 1}^K {\left\| {{{{\bm{\hat p}}}_{(m,k)}} - {{\bm{p}}_{(m,k)}}} \right\|} } 
	\label{eq:loss_function_loc}
	\end{equation}
	where ${\bm{\hat p}}_{(m,k)}$ and ${\bm{ p}}_{(m,k)}$ are respectively the predicted and true 2-D locations of the $m$th person at $k$th time step; ${\left\| {{{{\bm{\hat p}}}_{(m,k)}} - {{\bm{p}}_{(m,k)}}} \right\|}$ represents the euclidean distance between the locations of ${\bm{\hat p}}_{(m,k)}$ and ${\bm{ p}}_{(m,k)}$. When training the network for localization, we will also adopt the scheme of PIT.
	
	\subsubsection{Hyperparameters}
	
	The values of $D$ and $K$ of the localization network are the same as the ones of the above person counting network. The number of 1d-convolution kernel, i.e. $C$, is set to 32. The width of the BiLSTM layers in the stage for extraction is set to 512. The width of the fully connected layer in the stage for extraction is set to $32 \cdot M$. The width of the BiLSTM layers in the stage for single-person localization is set to 64. The width of the fully connected layer in the stage for single-person localization is set to $2 \cdot M$.
	
	\subsection{A Baseline Method}
	Besides the proposed method based on deep learning, the traditional methods of independent component analysis (ICA) and its extension of dependent component analysis (DCA) \cite{li2010dependent} also seem applicable for the task of PIR-based multi-person localization. The methods of ICA/DCA are a series of unsupervised learning methods which aim at decomposing a mixing signal into components related to different sources. Therefore, we can first utilize an ICA/DCA method to separate the DHF of the PIR sensors into components related to different persons. Then, the location of each person can be obtained by applying a PIR-based single-person localization method on each separated component.
	
	\begin{figure}[h]
		%\vspace{-1em}
		\centering
		\includegraphics[width=0.75\linewidth]{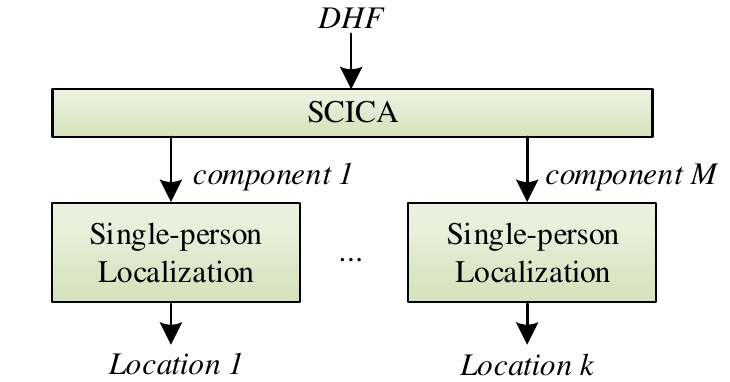}
		%\vspace{-1em}
		\caption{\bf Basic idea of utilizing SCICA for PIR-based localization.}%
		\label{fig:ica_loc_flow_chart}
		\vspace{-1em}
	\end{figure}
	
	In this subsection, we propose a baseline method based on the above idea. The flow chart of the proposed baseline method is illustrated in Fig. \ref{fig:ica_loc_flow_chart}. First, we utilize the method of SCICA \cite{davies2007source} to separate the DHF of PIR sensors into components corresponding to different persons. SCICA is an ICA/DCA method that aims at solving tasks of single channel-signal separation, just like the task of PIR signal separation. Before employing SCICA for separation, we need to model the DHF as a linear combination of a series of random variables:
	\begin{equation}
	{[...,x_t^{(n)},...,x_{t - D}^{(n)},...]^T} = {\bm{A}}{\bm{s}}
	\label{eq:ica_signal_model}
	\end{equation}
	where $x_t^{(n)}$ is the DHF of the $n$th PIR sensor at the sampling time $t$; $D$ is the sample number in each observation period; $\bm{A}$ is the mixing matrix; $\bm{s} \in \mathbb{R}^{N \cdot D}$ is a vector containing the random variables that compose the observed signal; N is the total number of PIR sensors. When executing the grouping step of SCICA, we also need to determine the number of moving persons in the observation period. To get it, we also adopt the person counting method proposed in Section \ref{sec:Network for person counting}. After executing SCICA and obtaining the separated components, we apply the PIR-based single-person localization method proposed in \cite{yang2019push} to estimate the location of each person.

	\section{Preprocessing and data augmentation}
	\label{sec:Preprocessing and data augmentation}
	\subsection{Preprocessing}
	\label{sec:preprocessing}

	Instead of directly utilizing the raw output of PIR sensors as the PIRNet's input, we perform two steps of preprocessing on it. The first step is denoising. In \cite{yang2019push}, the authors introduce that the raw output contains noise caused by the dynamic environmental factors (such as the wind), and the noise could degrade the localization accuracy. To alleviate the influence of the noise, they propose to suppress the noise through the inverse filter \cite{miyoshi1988inverse}. Therefore, we also utilize the inverse filter to process the raw output. Besides, the denoised raw output is essentially the DHF of the PIR sensor. The second preprocessing step is normalization. This step aims at removing the influence of changeable environmental temperature. In \cite{narayana2015pir}, the authors introduce that the amplitude of PIR sensors' output can be influenced by the environmental temperature. Specifically, the higher the temperature, the lower the amplitude. To improve the generalization ability of PIRNet in environments of different temperatures, we perform normalization on the denoised signal as follows: 
	\begin{equation}
	{\bm{x}} = {\bm{s}}/\sum\limits_{i = 1}^N {{\sigma _i}} 
	\label{eq:normalization scheme}
	\end{equation}
	where $\bm{s}$ is a matrix, each row of which contains the DHF of a PIR sensor in an observation period of $D$; $\sigma _i$ is the standard deviation of the $i$th row of $\bm{s}$.
	
	\subsection{Data augmentation}
	\label{sec:data augmentation}
	
	Data augmentation is a commonly used scheme to improve the performance of deep learning \cite{perez2017effectiveness}. Its basic idea is to increase the diversity of the training dataset through some prior knowledge of the task to be solved. For example, when training a neural network for image classification, people usually extends the training dataset by rotating and rescaling the initial images. The treatments of rotating and rescaling are based on the prior knowledge that the class of an image should not change even if the image is rotated and rescaled. Below, we will propose two data augmentation strategies for the training dataset of PIR-based localization.

	\subsubsection{Increasing diversity of moving speed}
	\label{sec:Stretching and Compressing}
\begin{figure}[h]
		\centering
		\includegraphics[width=0.9\linewidth]{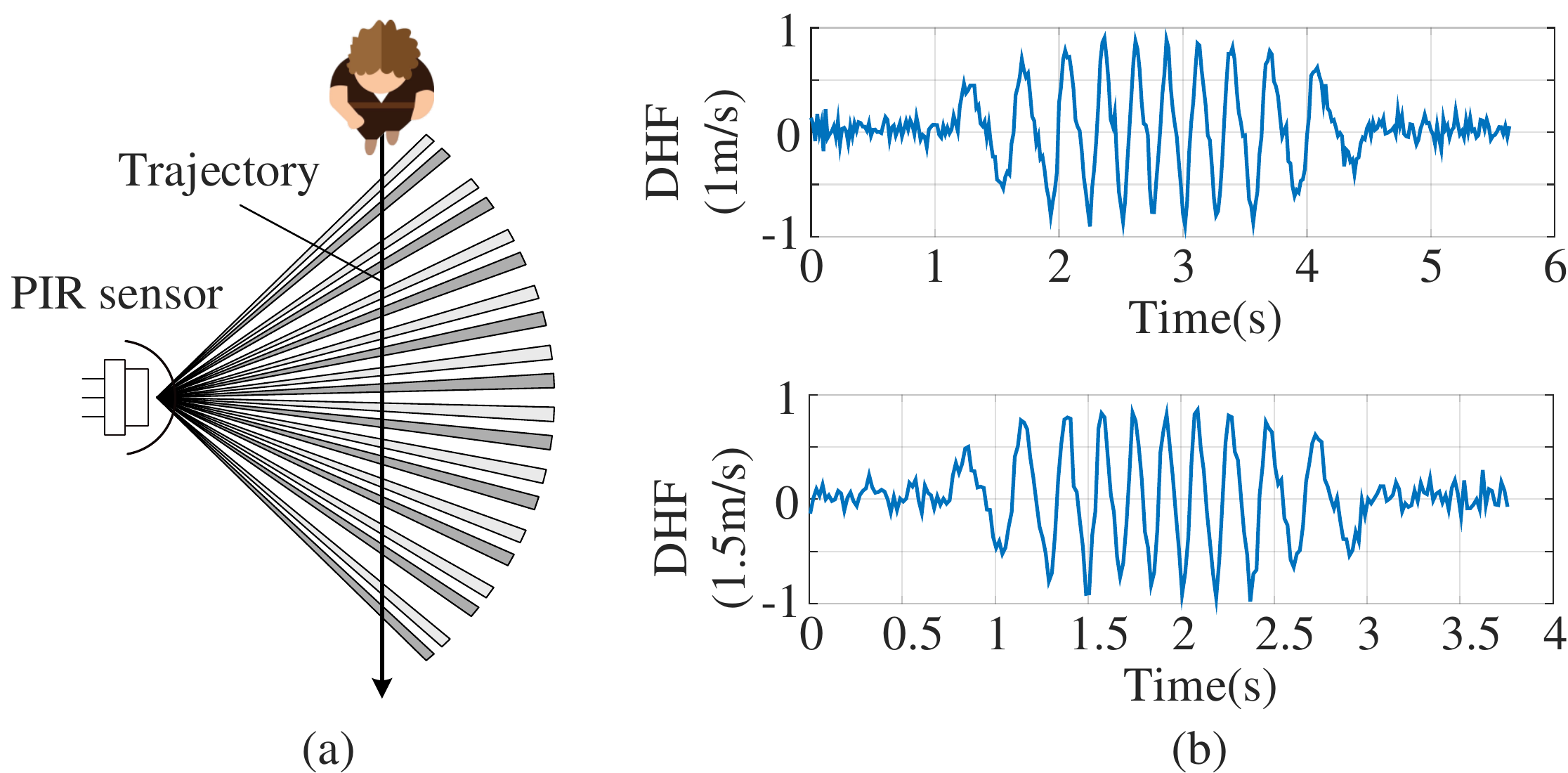}
		\caption{\bf DHFs in scenarios where a person moves along a trajectory with different speeds of 1m/s and 1.5m/s.}%
		\label{fig:same-trace-different-speed}
		%\vspace{-0.4em}
	\end{figure}

	The first data augmentation strategy aims at increasing the diversity of the training data from the aspect of the speed of moving persons. The prior knowledge behind the first strategy is that, when a person moves along the same trajectory with different speeds, a PIR sensor's DHFs (i.e. the inversely filtered raw output) will be of the same shape but different length. 
	For example, as shown in Fig. \ref{fig:same-trace-different-speed}(a), when a person moves in front of a PIR sensor at different speeds of 1m/s and 1.5m/s, the corresponding DHFs of the PIR sensor are shown in Fig. \ref{fig:same-trace-different-speed}(b).
	It can be seen that the DHF of high speed is approximately the same as the compressed signal at the low speed.
	
	Through the above prior knowledge, we propose to augment the training dataset by stretching or compressing the DHF of initially samples to simulate samples at lower or higher speed. The realization of stretching and compressing the inputs and labels for training is described as follows:
	\begin{equation}
	\left\{ \begin{array}{l}
	s_{DA}^{(i)}(t) = {s^{(i)}}(\frac{t}{a}) \vspace{1ex} \\
	x_{DA}^{(m)}(t) = {x^{(m)}}(\frac{t}{a}) \vspace{1ex} \\
	y_{DA}^{(m)}(t) = {y^{(m)}}(\frac{t}{a})
	\end{array} \right.
	\label{eq:data augmentation scheme}
	\end{equation}
	where  $s^{(i)}$ represents the DHF of the $i$th PIR sensor; $s_{DA}^{(i)}$ represents the stretched or compressed signal; $a$ is the parameter to control the degree of stretching or compressing; $x^{(m)}$ and $y^{(m)}$ are the ground-truth coordinates of the $m$th person corresponding corresponding to the initial signal; $x_{DA}^{(m)}$ and $y_{DA}^{(m)}$ is the ground-truth of the stretched or compressed signal. In the experiment, we respectively set $a$ to 1.2 and 0.8 to simulate the training samples of higher or lower speed.
	
	\subsubsection{Increasing diversity of background noise}
	\label{sec:Amplifying and Suppressing}
	The second data augmentation strategy aims at increasing the diversity of the training data from the aspect of background noise. The prior knowledge behind this strategy is that the background infrared radiation may distort the amplitude of the peaks/troughs of a PIR sensor's DHF. In the following, we first introduce why the background radiation could distort the DHF, and then propose the detailed process of this data augmentation strategy.
	
\begin{figure}[h]
		\centering
		\includegraphics[width=0.95\linewidth]{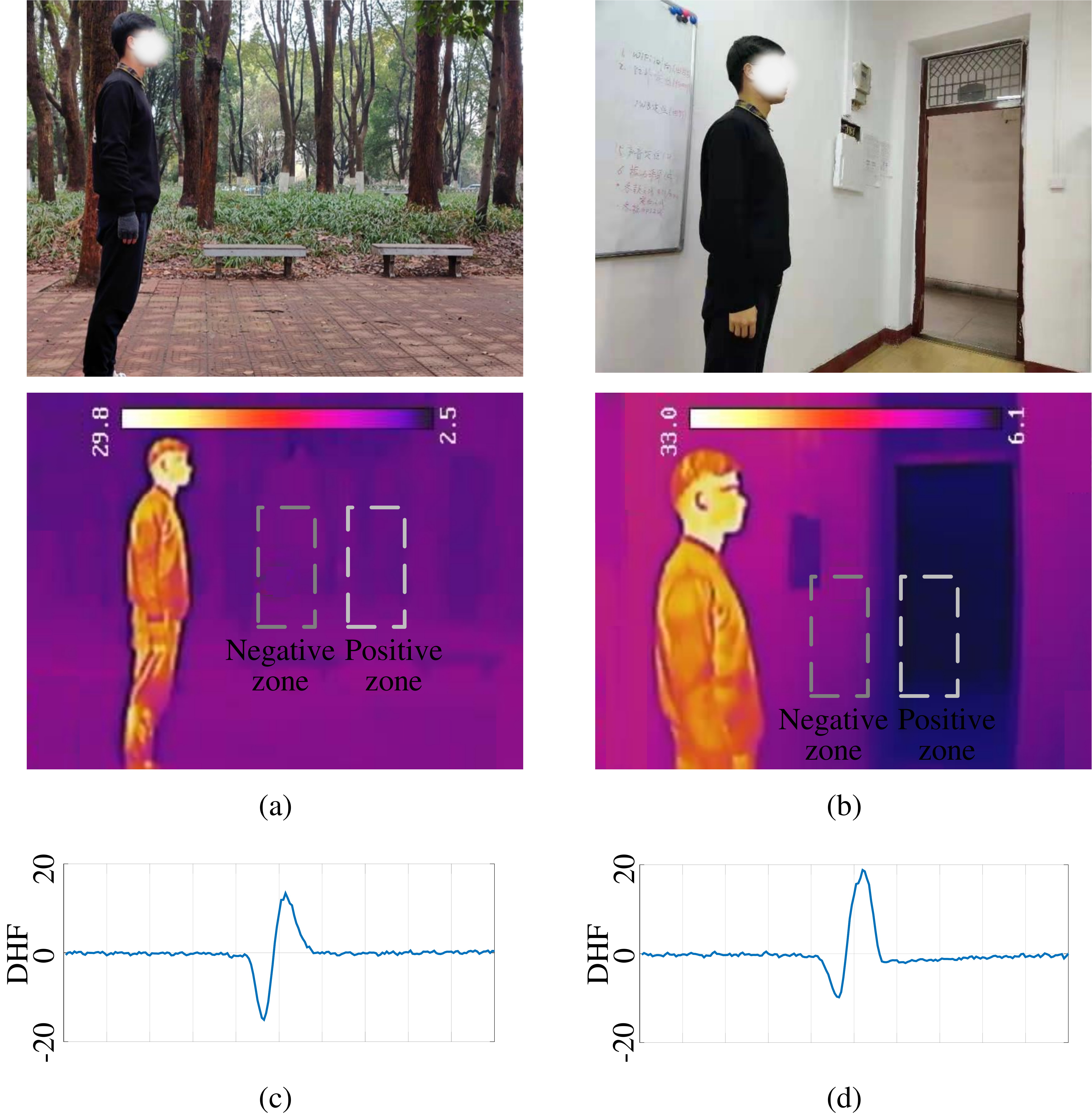}
		\vspace{-0.5em}
		\caption{\bf The DHFs of a same trajectory in different environments: (a) - (c) the visual and infrared images of different environments; (d) - (f) the corresponding DHFs when a person moves from left to right.}%
		
		\label{fig:distortion_of_background_radiation}
		%\vspace{-1.5em}
	\end{figure}

	Fig. \ref{fig:distortion_of_background_radiation} illustrates an example which shows how the DHF is distorted by the background radiation. In this example, a person successively moves across the negative and positive zones of a PIR sensor in two scenarios. The visual and infrared images of the scenarios are shown in Fig. \ref{fig:distortion_of_background_radiation}(a) and (b). The corresponding DHFs of the PIR sensor are shown in Fig. \ref{fig:distortion_of_background_radiation}(c) and (d). It can be seen that, in scenario (a), the amplitudes of the DHF's peak and trough are similar. However, in scenario (b), the amplitude of the peak becomes much higher than the amplitude of the trough.
	
\begin{figure}[h]
		%\vspace{-0.5em}
		\centering
		\includegraphics[width=0.6\linewidth]{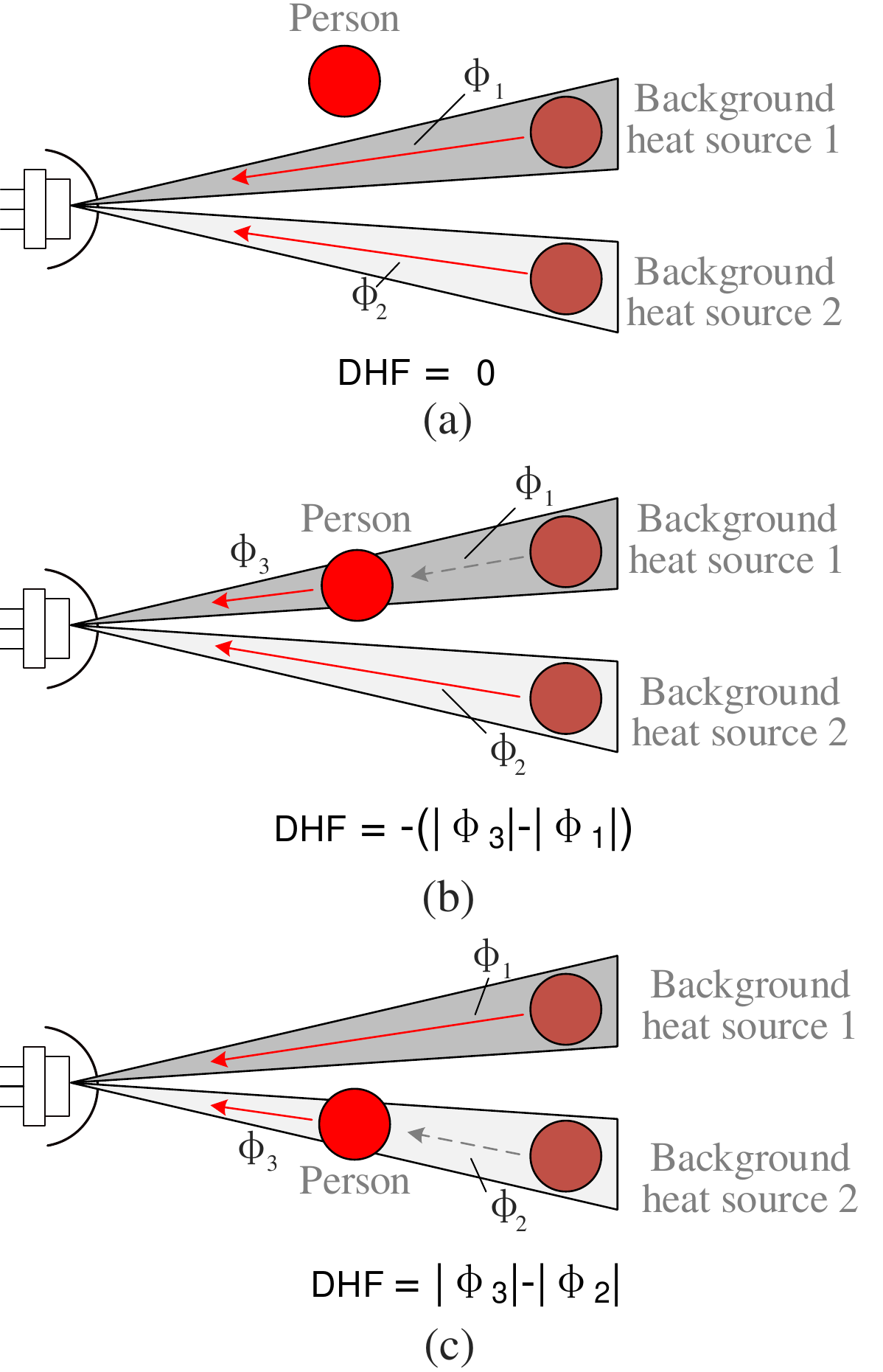}
		\vspace{-0.5em}
		\caption{\bf Influence of the background heat sources.}%
		
		\label{fig:reason_of_background_noise}
		%\vspace{-1.5em}
	\end{figure}

	Fig. \ref{fig:reason_of_background_noise} illustrates the reason behind the phenomenon shown in Fig. \ref{fig:distortion_of_background_radiation}. In Fig. \ref{fig:reason_of_background_noise}, we denote the DHFs caused by the background heat sources in negative and positive zones by $\phi_1$ and $\phi_2$, respectively, and denote $\phi_3$ as the DHF caused by the person. In practice, $\phi_1$ and $\phi_2$ are usually considered as constant since the background radiation seldomly changes. Therefore, $\phi_1$ and $\phi_2$ can be removed from the DHF by letting the DHF become a zero-mean signal. Then, as shown in Fig. \ref{fig:reason_of_background_noise}(a), when a person is outside the detection zone, the DHF of the PIR sensors is calculated as follows:
	\begin{equation}
	\phi_a = (|{\phi _2}| - |{\phi _1}|) - (|{\phi _2}| - |{\phi _1}|) = 0.
	\label{eq:DHF_a}
	\end{equation}
	When the person moves into the negative zone as shown in Fig. \ref{fig:reason_of_background_noise}(b), the background heat source in the negative zone will be occluded, and only the radiation of the person can achieve the sensor's negative element. In this situation, the sensor's DHF becomes:
	\begin{equation}
	\phi_b = (|{\phi _2}| - |{\phi _3}|) - (|{\phi _2}| - |{\phi _1}|) =  - (|{\phi _3}| - |{\phi _1}|).
	\label{eq:DHF_b}
	\end{equation}
	For a similar reason, when the person moves into the positive zone as shown in Fig. \ref{fig:reason_of_background_noise}(c), the sensor's DHF becomes:
	\begin{equation}
	\phi_c = (|{\phi _3}| - |{\phi _1}|) - (|{\phi _2}| - |{\phi _1}|) = |{\phi _3}| - |{\phi _2}|.
	\label{eq:DHF_c}
	\end{equation}
	
	From (\ref{eq:DHF_b}) and (\ref{eq:DHF_c}) we can see that, in the situation illustrated in Fig. \ref{fig:distortion_of_background_radiation}(a), where $|{\phi _1}| \approx |{\phi _2}|$, we can obtain that $|{\phi _b}| \approx |{\phi _c}|$. That is why the amplitudes of the peak and trough shown in Fig. \ref{fig:distortion_of_background_radiation}(c) are similar. In contrast, in the situation illustrated in Fig. \ref{fig:distortion_of_background_radiation}(b), where $|{\phi _1}| < |{\phi _2}| < |{\phi _3}|$, we can obtain that $|{\phi _b}| > |{\phi _c}|$. That is why the amplitude of the peak is higher than the amplitude of the trough as shown in Fig. \ref{fig:distortion_of_background_radiation}(d).
	
	Therefore, we can simulate training samples of different background distortion through enhancing/suppressing some peaks/troughs of the DHFs of initial training samples. Specifically, we first randomly select a series of peaks/troughs. We assume that the occurrence time of a selected peak/trough is $t_2$, and the occurrence times of its nearest adjacent two troughs/peaks are $t_1$ and $t_3$. Then, we enhance/suppress the selected peaks/troughs through the following formula:
	\begin{equation}
	%\small
	s_{DA}^{(i)}(t) = \left\{ \begin{array}{l}
	w \cdot ({s^{(i)}}(t) - {s^{(i)}}({t_1})) + {s^{(i)}}({t_1})\\
	,{t_1} \le t \le {t_2};\\
	\frac{{t - {t_2}}}{{{t_3} - {t_2}}} \cdot ({s^{(i)}}({t_3}) - s_{DA}^{(i)}({t_2})) + s_{DA}^{(i)}({t_2})\\
	,{t_2} < t \le {t_3}
	\end{array} \right.
	\label{eq:enhancing_or_suppressing}
	\end{equation}
	where $w$ is a random value from 0.5 to 1.5 which represents the degree of enhancing/suppressing; $s^{(i)}$ is the DHF of the $i$th PIR sensor before the enhancing/suppressing; $s_{DA}^{(i)}$ is the DHF after the enhancing/suppressing. In the experiment, we randomly enhance/suppress 10\% peaks/troughs of the initial training samples.

	\section{Evaluations}
	\label{sec:evaluations}
	
	\subsection{Setups}
	\label{sec:dataset}
	\subsubsection{Experimental environments}
	
\begin{figure}[h]
		\centering
		\includegraphics[width=1\linewidth]{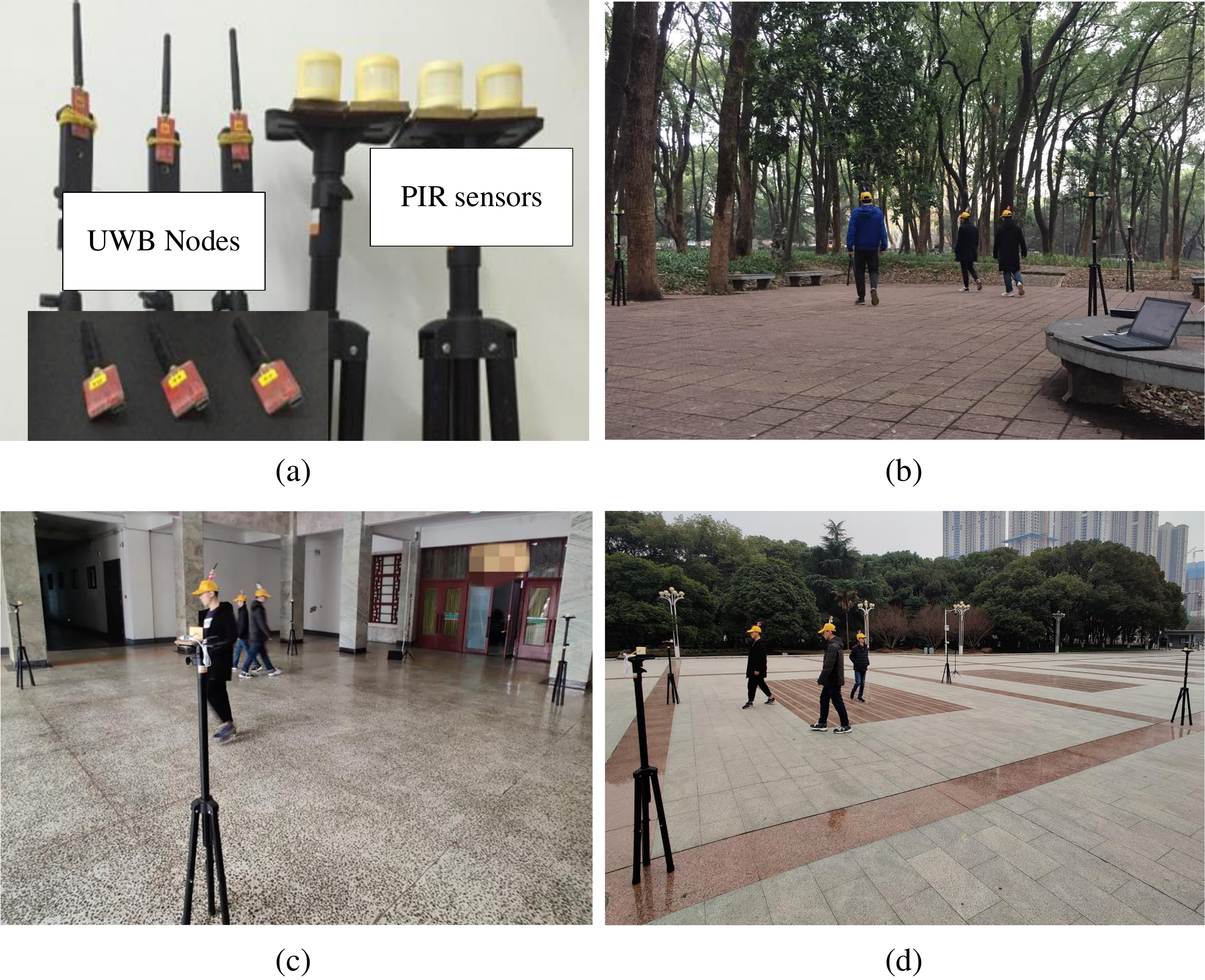}
		\caption{\bf Hardware platform and environments for data collection: (a) hardware platform; (b) training environment; (c)-(d) testing environment.}%
		\label{fig: real scenes}
		\vspace{-1.5em}
	\end{figure}

	We test our method in a $7m \times 7m$ area with 4 PIR sensors deployed on the 4 corners. As shown in Fig. \ref{fig: real scenes}(a), the PIR sensor and Fresnel lens array adopted in our experiment are respectively Tranesen-PCD-2F21 \cite{ManualOfPIR} and YUYING-8719 \cite{ManualOfLens}, which are both off-the-shelf and widely used in many PIR-based applications.
	In addtion, the ground-truth locations of multiple persons are derived through a UWB-based localization system YCHIOT-MINI3S \cite{ManualOfUwb}, whose average localization error is about 10cm. In the experiments, the client UWB nodes are fixed on the caps wore by each participant.
	
	The training data is collected in a park as shown in Fig. \ref{fig: real scenes}(b). For scenarios of 1, 2, and 3 persons, we respectively collect training data of 5 hours. The validation data of each scenario, which is used for the early stopping strategy in the training procedure, is also collected in the environment shown \ref{fig: real scenes}(b). The length of the validation data for each scenario is 15 minutes. On the other hand, the testing data is collected in two different environments, an indoor environment and an outdoor environment respectively shown in Fig. \ref{fig: real scenes}(c) and (d). The length of the testing data for each scenario is 30 minutes (15 minutes in indoor environment and 15 minutes in outdoor environment). In addition, the PIR sensors utilized for collecting the data sets for training and testing are different but of the same type.

	%\subsubsection{Environment for data collection}
	
	\subsubsection{Hyperparameters of the training procedure}
	\label{sec:training procedure}
	
	When training the PIRNet, we adopt the algorithm of Adam \cite{kingma2014adam}. The hyper-parameters of $\beta_1$, $\beta_2$ and $\varepsilon$ of Adam are respectively set to 0.9, 0.999 and ${10^{ - 8}}$. The learning rate is set to 0.001. To alleviate over-fitting, we adopt the scheme of early stopping in the training procedure. Specifically, the training procedure will be ended when the validation loss continuously increases in 5 epochs. The model used for testing is the one with the lowest validation loss in the training procedure.

	\subsection{Performance of PIRNet}
	\label{sec:Performance of PIRNet}
	
\begin{figure}[h]
		\centering
		\includegraphics[width=0.9\linewidth]{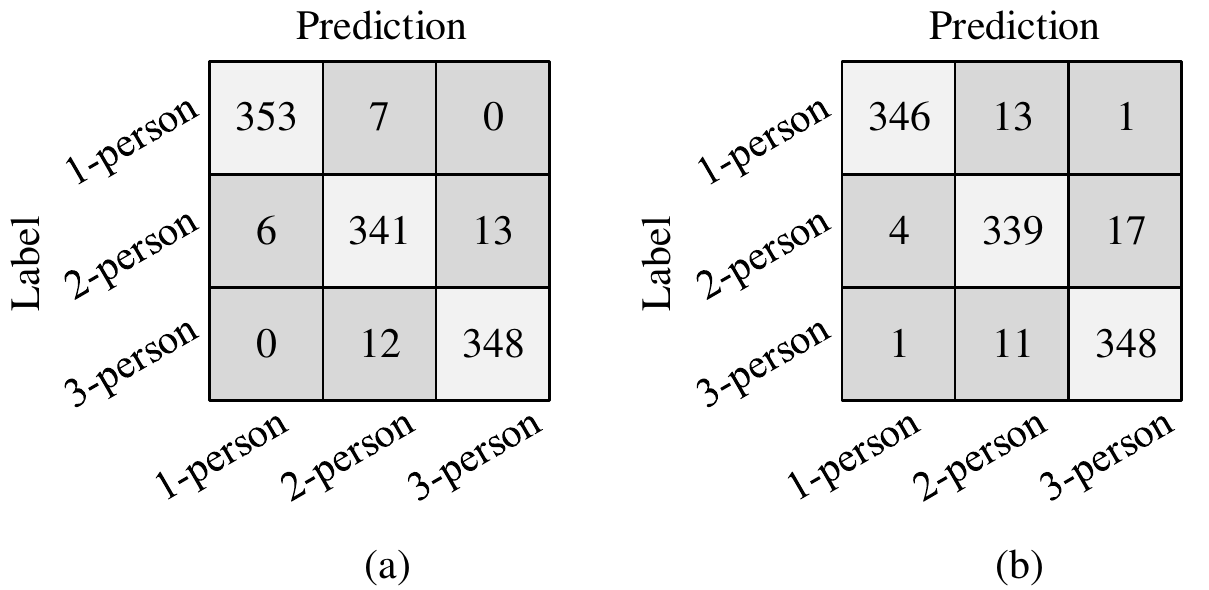}
		\caption{\bf Confusion matrixes of the person counting network in indoor and outdoor testing environments: (a) indoor; (b) outdoor.}%
		\label{fig:confusion matrix}
		%\vspace{-1em}
		% 绘图 Fig34_PIRNet_loc_error_cdf_v2.m
	\end{figure}

\begin{figure}[h]
		\centering
		\includegraphics[width=0.65\linewidth]{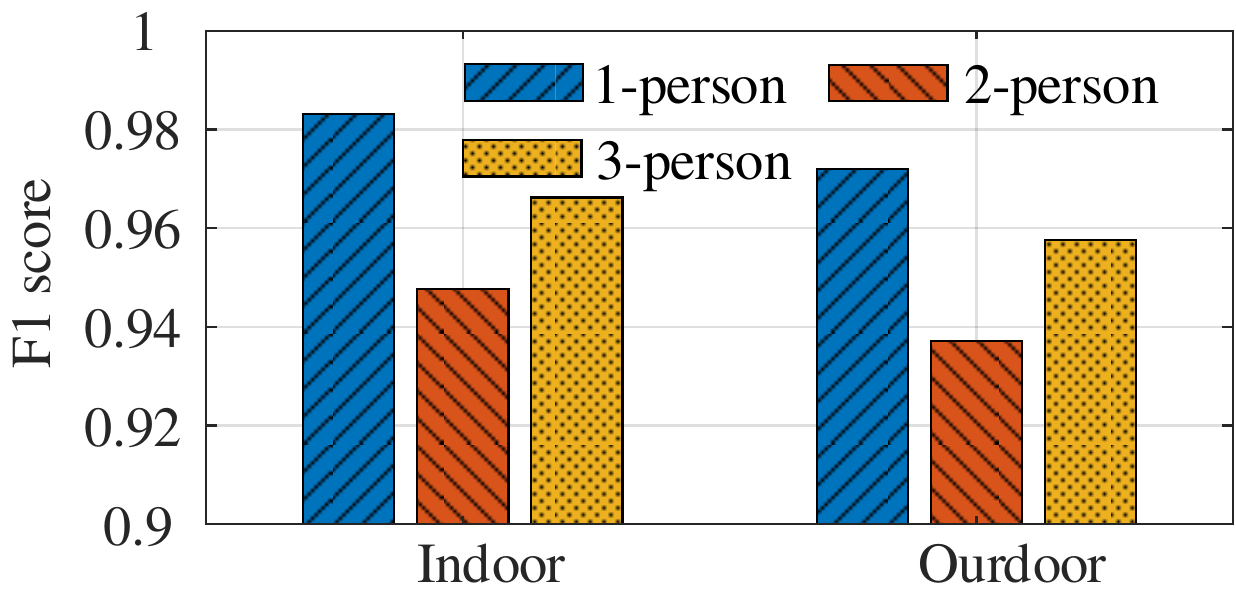}
		\caption{\bf F1 scores of 1-person, 2-person, and 3-person scenarios in indoor and outdoor testing environments.}%
		\label{fig:F1 scores}
		%\vspace{-0.4em}
		% 绘图 Fig34_PIRNet_loc_error_cdf_v2.m
	\end{figure}

	In this section, we will demonstrate the PIRNet's accuracies of person counting and M-person localization. Fig. \ref{fig:confusion matrix} illustrates the confusion matrixes of the person counting task in the indoor and outdoor testing environments. The accuracy in the indoor and outdoor environments is 96.4\% and 95.6\%, respectively. It can be seen that the counting accuracy of the indoor environment is slightly higher than the outdoor environment. The reason is that the outdoor environment is uncontrolled and contains more noise sources, such as the wind and moving persons outside the supervision area. In addition, we also illustrate the F1 scores of 1-person, 2-person, and 3-person scenarios in Fig. \ref{fig:F1 scores}. It can be seen that the F1 score of the 2-person scenario is slightly lower than the other two scenarios. The reason is that the 2-person samples are more confusing than the 1-person and 3-person samples. Specifically, as shown in Fig. \ref{fig:confusion matrix}, the samples of 2-person are possible be wrongly classified as samples of 1-person and 3-person, but the samples of 1-person and 3-person are rarely wrongly classified as each other.
	
\begin{figure}[h]
		\centering
		\includegraphics[width=0.85\linewidth]{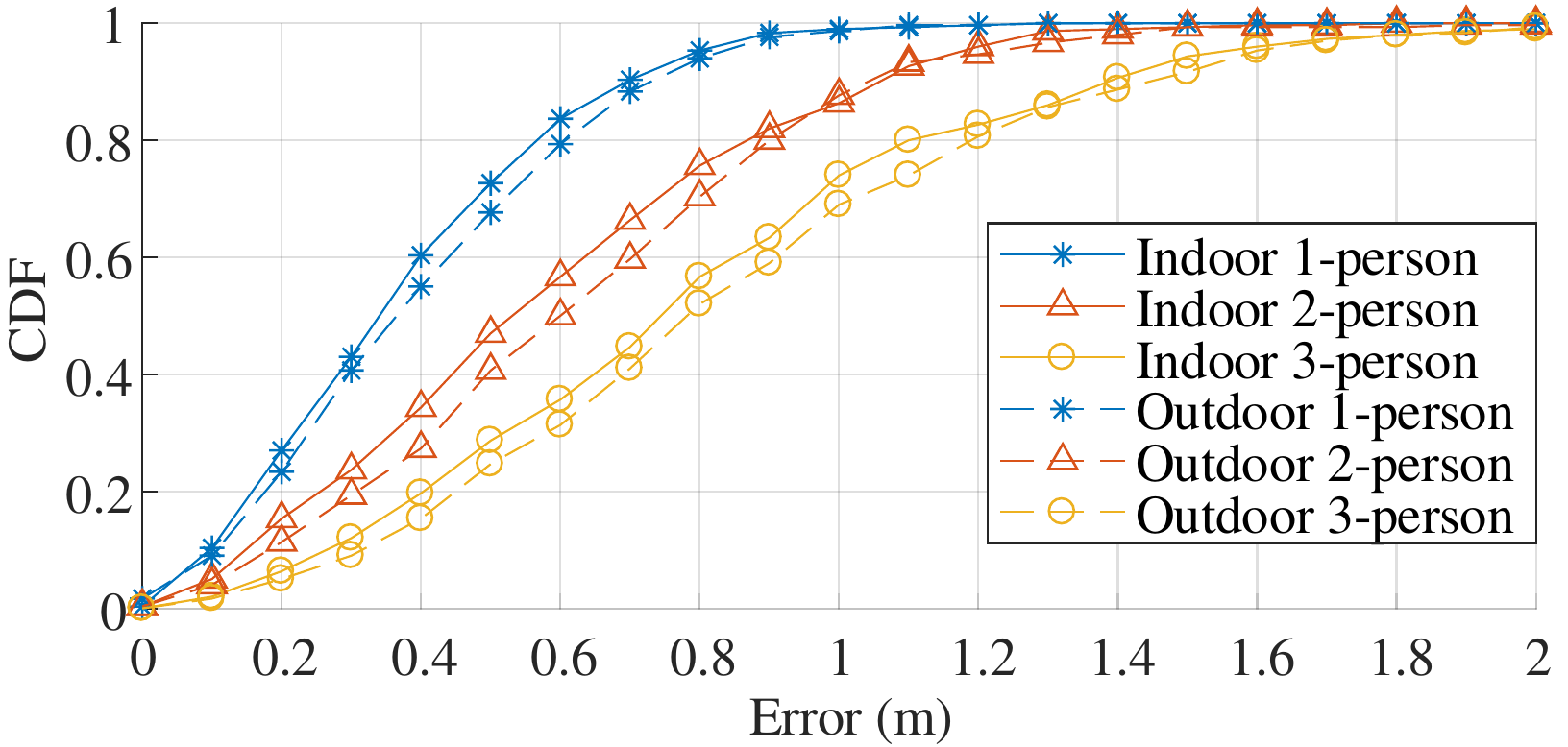}
		\caption{\bf CDFs of localization errors in the indoor and outdoor testing environments.}%
		\label{fig: indoor outdoor localization error}
		\vspace{-1em}
		% 绘图 Fig34_PIRNet_counting_accuracy.m
	\end{figure}

	In Fig. \ref{fig: indoor outdoor localization error}, we illustrate the cumulative distribution function (CDF) of localization error in the scenarios of 1, 2, and 3 persons. In the outdoor environment, the average errors of 1-person, 2-person, and 3-person are respectively 0.43m, 0.65m, and 0.84m. In the indoor environment, the average errors in scenarios of 1, 2, and 3 persons are 0.41m, 0.59m, and 0.80m, respectively. It can be seen that the localization accuracy in the controlled indoor testing environment is also slightly better than the uncontrolled outdoor testing environment.
	
	\subsection{Influence of training data length}
	In this section, we demonstrate the influence of training data length to the testing performance. For brevity, the following testing experiments no longer distinguish between indoor and outdoor environments. In Table \ref{tab:counting accuracy and training data length}, we illustrate the F1-scores of scenario of 1, 2, and 3 persons. It can be seen that, utilizing 3 hours of training data, the testing F1-scores of all scenarios are all higher than 90\%.
	
	\renewcommand\arraystretch{1.3}
	\begin{table}[h]
		\caption{\bf The influence of training data length to F1-scores of person counting.}
		\vspace{-0.5em}
		\label{tab:counting accuracy and training data length}
		\centering
		%\small
		\begin{tabular}{|c|*{5}{c|}}\hline
			%\backslashbox{Error}{Scenario}
			{\backslashbox{F1-score (\%)\kern-1em}{\kern-1em Length (h)}} &\makebox[1em]{1} &\makebox[1em]{2} &\makebox[1em]{3} &\makebox[1em]{4} &\makebox[1em]{5}\\ \hline
			1-person & 79.1 & 89.1 & 93.6 & 94.7 & \bf{97.7}\\ \hline
			2-person & 71.1 & 82.8 & 90.4 & 92.2 & \bf{94.2}\\ \hline
			3-person & 75.8 & 85.8 & 91.6 & 91.7 & \bf{96.2}\\ \hline
		\end{tabular}
		\vspace{-1em}
		
		% 原始数据: Tab1_f1_score_of_different_training_length.m
	\end{table}
	
	In Table \ref{tab:localization accuracy and training data length}, we illustrate the relationship between the testing localization errors and training data length. It can be seen that, to achieve the average localization error lower than 1m, the scenarios of 1, 2, and 3 persons respectively require training data of 1 hour, 2 hours, and 4 hours. 
	
	\renewcommand\arraystretch{1.3}
	\begin{table}[h]
		\centering
		\caption{\bf The influence of training data length to the mean value and standard deviation of the absolute  localization error.}
		\vspace{-0.5em}
		\label{tab:localization accuracy and training data length}
		%\small
		\begin{tabular}{|c|*{6}{c|}}\hline
			%\backslashbox{Error}{Scenario}
			\multicolumn{2}{|c|}{\backslashbox{Error (m)\kern-1em}{Length (h)}} &\makebox[1em]{1} &\makebox[1em]{2} &\makebox[1em]{3} &\makebox[1em]{4} &\makebox[1em]{5}\\ \hline
			\multirow{2}{30pt}{1-person} 
			& Mean & 0.77 & 0.61 & 0.52 & 0.47 & \bf{0.42}\\ \cline{2-7}
			& Std & 0.41 & 0.31 & 0.29 & 0.24 & \bf{0.22}\\ \hline
			
			\multirow{2}{30pt}{2-person}
			& Mean & 1.02 & 0.94 & 0.82 & 0.71 & \bf{0.62}\\ \cline{2-7}
			& Std & 0.55 & 0.47 & 0.46 & 0.36 & \bf{0.32}\\ \hline
			\multirow{2}{30pt}{3-person} 
			& Mean & 1.25 & 1.16 & 1.08 & 0.92 & \bf{0.82}\\ \cline{2-7}
			& Std & 0.66 & 0.60 & 0.54 & 0.46 & \bf{0.41}\\ \hline
		\end{tabular}
		\vspace{-1em}
		
		%原始数据: ./Tab3_loc_error_of_different_traning_data_length.m
	\end{table}
	
	\subsection{Influence of sensors number}
\begin{figure}[h]
		\vspace{-0.5em}
		\centering
		\includegraphics[width=0.45\linewidth]{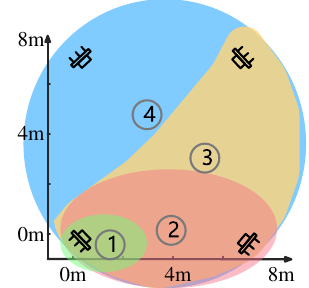}
		\caption{\bf Deployments of scenarios when utilizing 1, 2, 3, and 4 PIR sensors.}%
		\label{fig: Deployments of 1234 PIR}
		\vspace{-1em}
	\end{figure}

	In this section, we validate the relationship between the testing performance and the utilized number of PIR sensors. The experiments are conducted in 4 scenarios where there are 1, 2, 3, and 4 PIR sensors deployed in the $7m \times 7m$ environment as shown in Fig. \ref{fig: Deployments of 1234 PIR}. Specifically, in the first scenario, one PIR sensor is deployed at $(0m,0m)$. In the second scenario, two PIR sensors are deployed at $(0m,0m)$ and $(7m,0m)$. In the third scenario, three PIR sensors are deployed at $(0m,0m)$, $(7m,0m)$ and $(7m,7m)$. In the fourth scenario, four PIR sensors are deployed at $(0m,0m)$, $(7m,0m)$, $(7m,7m)$, and $(0m,7m)$, respectively. In each scenario, the utilized training data length for each scenario is 5 hours.
	
	\renewcommand\arraystretch{1.3}
	\begin{table}[h]
		\centering
		\caption{\bf The influence of sensors number to F1-scores of person counting.}
		\vspace{-0.5em}
		\label{tab:counting accuracy and sensors number}
		%\small
		\begin{tabular}{|c|*{4}{c|}}\hline
			%\backslashbox{Error}{Scenario}
			{\backslashbox{F1-score (\%)\kern-1em}{Number}} &\makebox[1em]{1} &\makebox[1em]{2} &\makebox[1em]{3} &\makebox[1em]{4} \\ \hline
			1-person & 95.3 & 96.6 & 97.2 & \bf{97.7} \\ \hline
			2-person & 91.1 & 92.7 & 94.2 & \bf{94.2} \\ \hline
			3-person & 93.3 & 93.8 & 96.2 & \bf{96.2} \\ \hline
		\end{tabular}
		
		%原始数据: ./Tab2_f1_score_of_pirnum_v2.m
		\vspace{-0.5em}
	\end{table}

	Table \ref{tab:counting accuracy and sensors number} illustrates the average F1-scores of each scenario. It can be seen that, even utilizing 1 PIR sensor, the F1-scores of all scenario are higher than 90\%. In Table \ref{tab:localization accuracy and sensors number}, we demonstrate the testing localization errors when utilizing different number of PIR sensors. It can be seen that, with 5 hours training data, the scenarios of 1, 2, and 3 persons respectively require at least 2, 3 and 4 sensors to achieve the average localization error lower than 1m.

	\renewcommand\arraystretch{1.3}
	\begin{table}[h]
		\centering
		\caption{\bf The influence of sensors number to the mean value and standard deviation of the absolute localization error.}
		\vspace{-0.5em}
		\label{tab:localization accuracy and sensors number}
	
		\begin{tabular}{|c|*{5}{c|}}\hline
			%\backslashbox{Error}{Scenario}
			\multicolumn{2}{|c|}{\backslashbox{Error (m)\kern-1em}{Number}} &\makebox[1em]{1} &\makebox[1em]{2} &\makebox[1em]{3} &\makebox[1em]{4} \\ \hline
			\multirow{2}{30pt}{1-person} 
			& Mean & 1.26 & 0.85 & 0.60 & \bf{0.42}\\ \cline{2-6}
			& Std & 0.67 & 0.44 & 0.32 & \bf{0.22}\\ \hline
			
			\multirow{2}{30pt}{2-person}
			& Mean & 1.49 & 1.13 & 0.81 & \bf{0.62}\\ \cline{2-6}
			& Std & 0.81 & 0.57 & 0.45 & \bf{0.32}\\ \hline
			\multirow{2}{30pt}{3-person} 
			& Mean & 1.90 & 1.51 & 1.16 & \bf{0.82}\\ \cline{2-6}
			& Std & 0.99 & 0.78 & 0.58 & \bf{0.41}\\ \hline
		\end{tabular}
		
		%原始数据：./Tab5_loc_error_of_different_pirNum_v2.m
		\vspace{-0.5em}
	\end{table}
	
	\subsection{Improvement of preprocessing and data augmentation}
	\label{sec:Improvement through preprocessing and data augmentation}
	In this section, we demonstrate the improvement brought by the techniques of preprocessing and data augmentation. From Table \ref{tab:preprocessing_data_augmentation}, it can be seen that only utlizing the prepocessing can improve the person counting accuracy by 0.9\%, and decrease the average localization error by $4.6cm$. In addition with the data augmentation, the person counting accuracy can be further improved by 0.5\%, and the average localization error can be further decreased by $2.7cm$. 
	
	\begin{table}[h]
		\scriptsize
		\centering
		\caption{\bf Improvement through the strategies of preprocessing and data augmentation: `PP' represents the preprocessing; `DA' represents the data augmentation.}
		\vspace{-0.5em}
		\label{tab:preprocessing_data_augmentation}
		
		\begin{tabular}{|c|c|c|c|c|}  
			\hline  
			\multirow{2}{30pt}{\centering Network} 
			& \multirow{2}{30pt}{\centering Accuracy (\%)}  
			& \multicolumn{3}{c|}{Mean / Std of Error (m)} \\
			\cline{3-5}
			& & 1 person& 2 persons& 3 persons\\
			\hline
			
			PIRNet & 94.2 & 0.48 / 0.25 & 0.72 / 0.39 & 0.96 / 0.50
			\\ \hline
			PIRNet+PP & 95.1 & 0.44 / 0.23 & 0.67 / 0.34 & 0.91 / 0.47
			\\ \hline
			PIRNet+PP+DA & \bf{96.1} & \bf{0.43 / 0.23} & \bf{0.62 / 0.32} & \bf{0.82 / 0.41}
			\\ \hline
		\end{tabular}  
		\vspace{-1em}
		
		% ./Tab4_loc_error_of_noPreprocess_onlyPreprocess.m
	\end{table}
	
	\subsection{Comparison with the baseline method of SCICA}
	
\begin{figure}[h]
		\centering
		\includegraphics[width=0.75\linewidth]{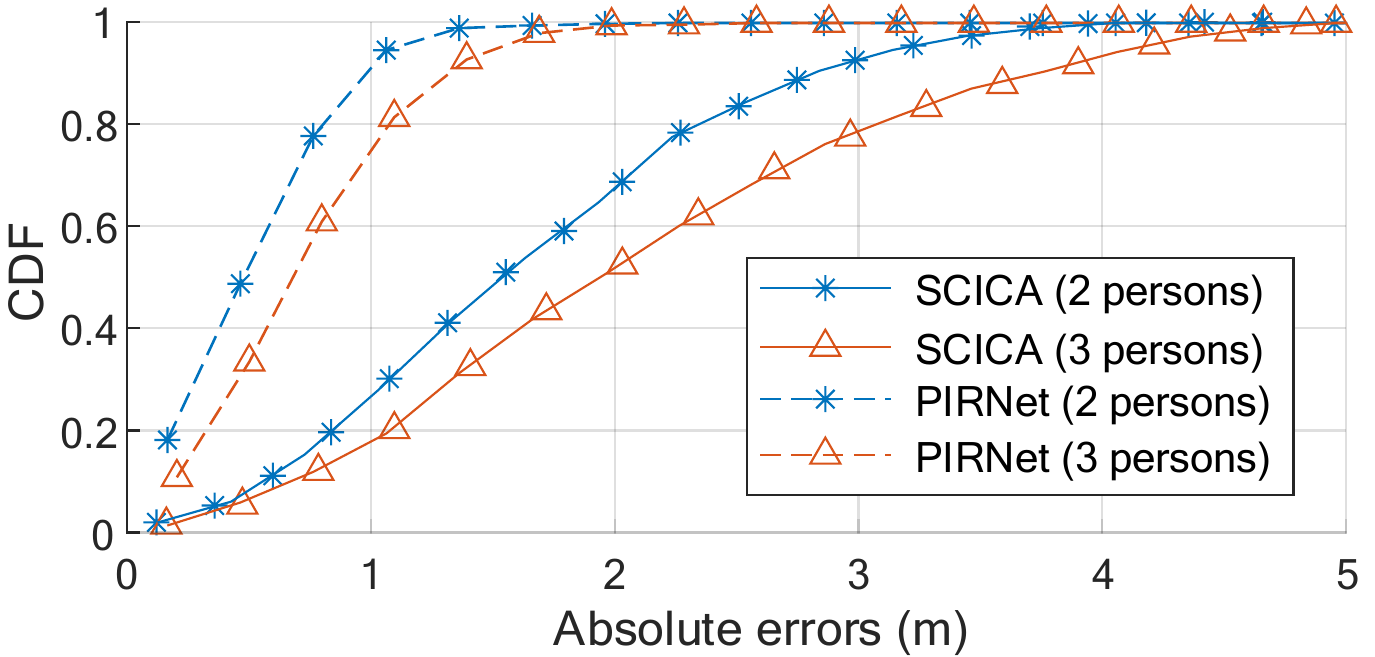}
		%\vspace{-1.5em}
		\caption{\bf CDFs of the absolute localization error respectively achieved by the SCICA method and PIRNet.}%
		\label{fig:cmp_mica}
		%\vspace{-1.5em}
	\end{figure}

	The data sets utilized for training and testing the baseline method based on SCICA are the same ones introduced in Section \ref{sec:dataset}. Fig. \ref{fig:cmp_mica} illustrates the comparison of the CDFs of the absolute localization errors of the SCICA-based method and PIRNet. The average absolute localization errors of the SCICA-based method are 1.76$m$ and 2.24$m$ in the scenarios of 2 and 3 persons, respectively. It can be seen that, the localization error of the baseline method is much higher than PIRNet. We believe the reason is that the DHFs corresponding to different persons have overlapped spectra. In \cite{davies2007source}, the authors introduce that the SCICA method could not well handle the independent signals that have substantially overlapping spectra.

	\subsection{Comparison with traditional PIR-based methods}
	\label{sec:Comparison with other PIR-based methods}
\begin{figure}[h]
		\centering
		\includegraphics[width=1\linewidth]{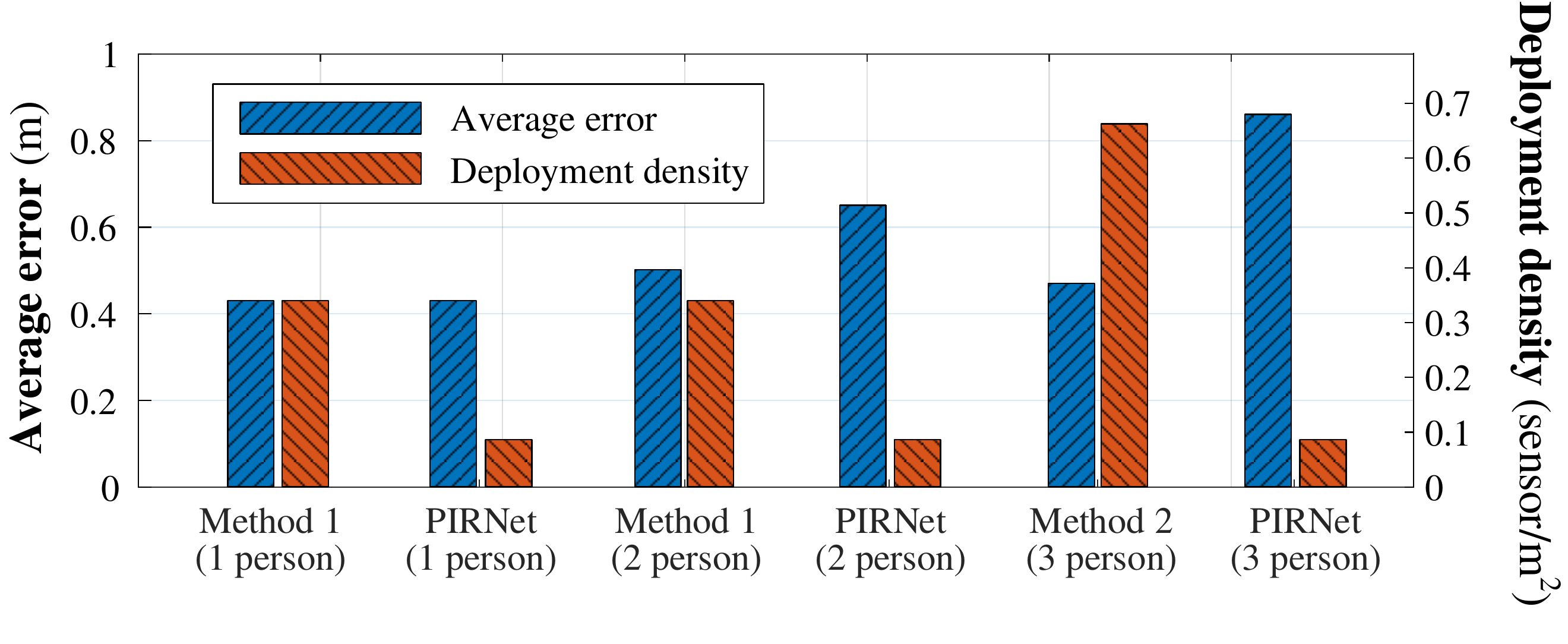}
		\vspace{-2em}
		\caption{\bf The average absolute localization error and deployment density of PIRNet and existing PIR-based methods of the highest localization accuracy.}%
		\label{fig:Comparing_with_traditional_method}
		%\vspace{-1.5em}
		% 绘图 Fig23_compare_PIRNet_with_other_methods.m
	\end{figure}

	% Compare the accuracy with other works
	% PIRNet与传统方法的对比
	In this section, we compared the average localization error and the deployment density of our system with some other PIR-based methods which achieve the state-of-the-art localization accuracy. To the best of our knowledge, for the scenarios of 1 and 2 person, the highest localization accuracy is achieved by the method proposed in \cite{yang2017multiple}. For the scenario of 3 persons, the highest localization accuracy is achieved by the method proposed in \cite{lu2016preprocessing}. We compare PIRNet and these two methods in Fig. \ref{fig:Comparing_with_traditional_method}. `Method 1' and `Method 2' in Fig. \ref{fig:Comparing_with_traditional_method} respectively refer to the methods proposed in \cite{yang2017multiple} and \cite{lu2016preprocessing}. For scenarios of 1 and 2 persons, the localization error of \cite{yang2017multiple} are respectively 0.43m and 0.50m, and the deployment density is about 0.34 $sensor/m^2$. For the scenario of 3 persons, the average localization error of \cite{lu2016preprocessing} is 0.47m, and the deployment density is about 0.67 $sensor/m^2$. It can be seen that, although the average localization error of PIRNet is slightly higher than the traditional methods, the PIRNet's deployment density is much lower than them, which is about 0.08 $sensor/m^2$.
	
	\subsection{Performance in noisy environments}
	
\begin{figure}[h]
		\vspace{-0.5em}
		\centering
		\includegraphics[width=0.55\linewidth]{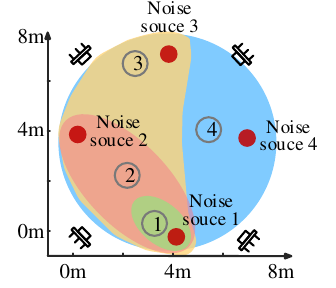}
		\caption{\bf Deployments of the noise sources in the noisy testing environment.}%
		\label{fig: Deployments of noise source}
		%\vspace{-1em}
	\end{figure}

	In section \ref{sec:Amplifying and Suppressing}, we introduced that the DHF of a PIR sensor would be distorted when there are background heat sources whose radiation intensities are much different from other surrounding objects. In the following, we specially test the influence of the above distortion on the performance of PIRNet. Since the radiation intensity of a person is usually evidently higher than other environmental objects, we let several persons stand around the testing environment as noise sources. In addition, the experiment is conducted in 4 scenarios where the numbers of noise sources are 1, 2, 3 and 4, respectively. Fig. \ref{fig: Deployments of noise source} illustrates the deployments of the noise sources in these scenarios. For each scenario, we collect 5 minutes of testing data when there are 1, 2, and 3 persons respectively. The testing environment for data collection is the same to one shown in Fig. \ref{fig: real scenes}(c).
	
	Table \ref{tab:performance_in_noisy_environment} illustrates the person counting accuracy and localization error in the above scenarios. It can be seen that, even when there are four noise sources in the testing environment, the person counting accuracy is still higher than 90\%, and the average localization error is still lower than 1m. 
	
	\begin{table}[h]
		\centering
		\caption{\bf The person counting accuracy and absolute localization error of PIRNet in the noisy testing environment.}
		\vspace{-0.5em}
		\label{tab:performance_in_noisy_environment}
		
		\begin{tabular}{|c|c|c|c|c|}  
			\hline  
			\multirow{2}{30pt}{\centering Noise sources}
			& \multirow{2}{30pt}{\centering Accuracy (\%)}  
			& \multicolumn{3}{c|}{Mean / Std of Error (m)} \\
			\cline{3-5}
			& & 1 person& 2 persons& 3 persons\\
			\hline
			
			0 & \bf{96.4} & \bf{0.41 / 0.22} & \bf{0.59 / 0.32} & \bf{0.79 / 0.40}
			\\ \hline
			1 & 95.3 & 0.43 / 0.21 & 0.62 / 0.31 & 0.80 / 0.39
			\\ \hline
			2 & 94.6 & 0.48 / 0.26 & 0.68 / 0.36 & 0.87 / 0.44
			\\ \hline
			3 & 92.2 & 0.51 / 0.25 & 0.74 / 0.38 & 0.91 / 0.47
			\\ \hline
			4 & 90.6 & 0.57 / 0.29 & 0.79 / 0.43 & 0.97 / 0.53
			\\ \hline
		\end{tabular}  
		\vspace{-1em}
		
		% ./Tab6_loc_error_of_noisy_enviroment.m
	\end{table}
	
	The robustness of PIRNet to background noise sources is partially contributed by the data augmentation strategy proposed in Section \ref{sec:Amplifying and Suppressing}, which aims at increasing the diversity of the training samples from the aspect of background noise. To demonstrate the effectiveness of this data augmentation strategy, in Table \ref{tab:performance_in_noisy_environment_no_DA}, we illustrate the performance of PIRNet which does not utilize this strategy in the training procedure. It can be seen that, without this strategy, the person counting accuracy decreases 1.6 percentage points on average, and the localization error increases 6.8$cm$ on average.
	
	\begin{table}[h]
		\centering
		\caption{\bf The person counting accuracy and absolute localization error of PIRNet in the noisy testing environment, without utilizing the data augmentation strategy for increasing the diversity of background noise.}
		\vspace{-0.5em}
		\label{tab:performance_in_noisy_environment_no_DA}
		
		\begin{tabular}{|c|c|c|c|c|}  
			\hline  
			\multirow{2}{30pt}{\centering Noise sources}
			& \multirow{2}{30pt}{\centering Accuracy (\%)}  
			& \multicolumn{3}{c|}{Mean / Std of Error (m)} \\
			\cline{3-5}
			& & 1 person& 2 persons& 3 persons\\
			\hline
			
			0 & \bf{96.0} & \bf{0.42 / 0.22} & \bf{0.62 / 0.33} & \bf{0.83 / 0.42}
			\\ \hline
			1 & 94.6 & 0.49 / 0.25 & 0.69 / 0.35 & 0.85 / 0.41
			\\ \hline
			2 & 93.0 & 0.54 / 0.30 & 0.75 / 0.40 & 0.98 / 0.49
			\\ \hline
			3 & 89.4 & 0.62 / 0.31 & 0.83 / 0.43 & 1.01 / 0.52
			\\ \hline
			4 & 88.1 & 0.63 / 0.32 & 0.86 / 0.47 & 1.07 / 0.58
			\\ \hline
		\end{tabular}  
		\vspace{-1em}
		
		% ./Tab7_loc_error_of_noisy_enviroment.m
	\end{table}
	
	\section{Conclusion and Future work}
	\label{sec:conclusion}
	In this paper, we propose a new method for PIR-based multi-person localization. The method is based on a deep neural network PIRNet which is designed through integrating the domain knowledge. In addition, we propose to perform the preprocessing and data augmentation to further improve the PIRNet's performance. Through the proposed method, we achieve similar localization accuracy compared with the state-of-art PIR-based methods, but with much lower deployment density. However, there are still two interesting problems for further exploration.

	One problem is how to improve the robustness of the proposed model in the environment of objects whose radiation intensities are evidently different from other surrounding objects. One direct way to solve this problem is to fine-tune the trained model by new training data collection in abundant noisy environments. Another possible way is to utilize the technique of adversarial training \cite{shaham2018understanding} to improve the generalization ability of the trained network.

	Another problem is how to reuse the training data for different deployment strategy. For example, in our experiment, the PIRNet is trained in a $7m \times 7m$ area where 4 PIR sensors are deployed at its 4 corners. However, this model can not be directly used in the environment of a different deployment of PIR sensors, e.g. a $4m \times 4m$ area where 4 PIR sensors deployed at the corners. 
	
\begin{figure}[h]
		\centering
		\includegraphics[width=0.8\linewidth]{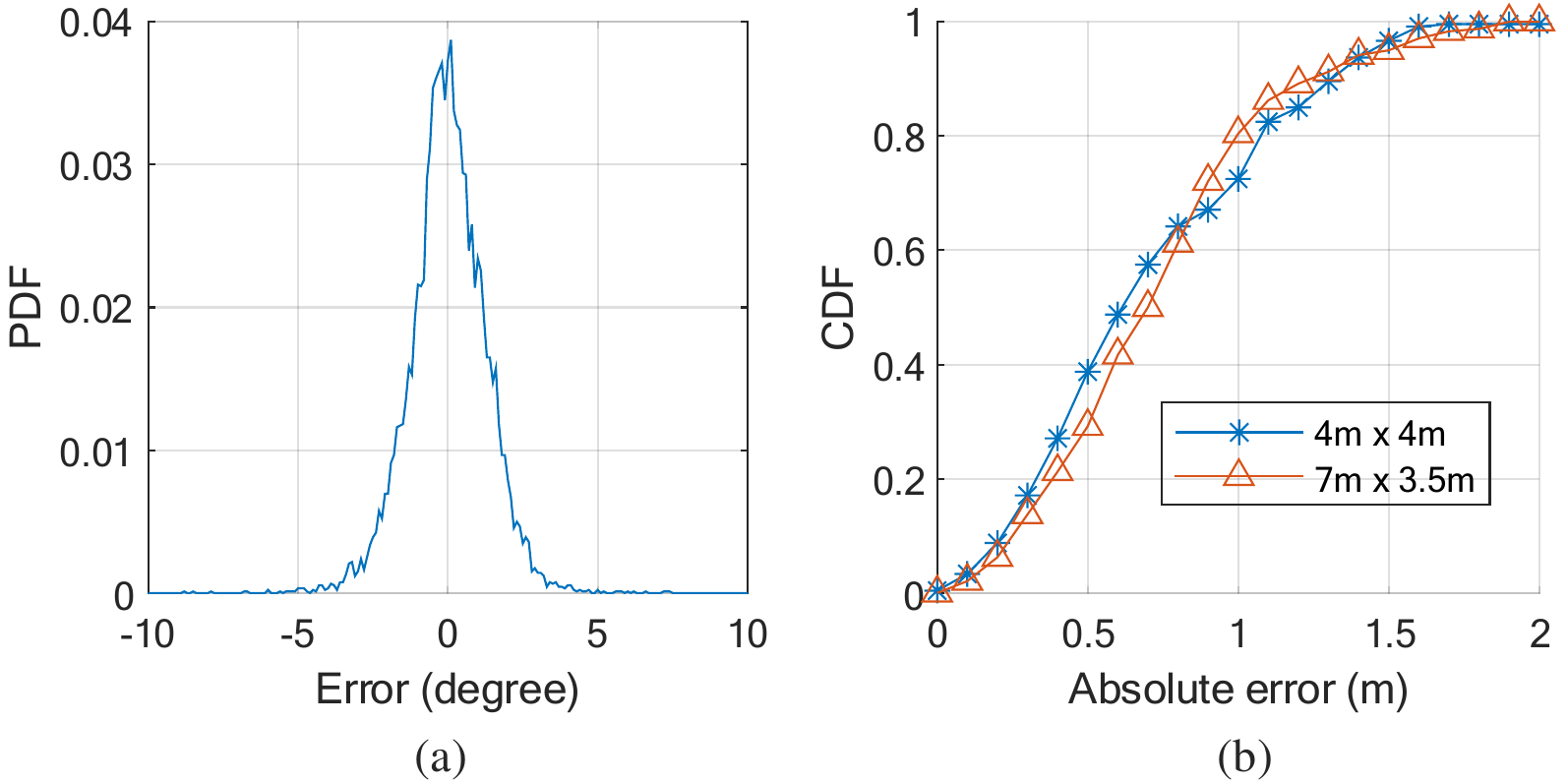}
		%\vspace{-2em}
		\caption{\bf Preliminary results of the method for reusing the training data for different deployment strategies.}%
		\label{fig:result_of_DeepPirates}
		%\vspace{-1.5em}
	\end{figure}

	A possible solution is that, instead of using an end-to-end deep learning model that directly generates locations, we can first train a deep learning model that generates a deployment-insensitive estimation, i.e. the azimuth change. Then the estimated azimuth changes are utilized to give locations of persons based on the non-data-driven technique proposed in \cite{yang2019push}. Based on this idea, we have achieved a preliminary and promising result which is demonstrated in Fig. \ref{fig:result_of_DeepPirates}. Specifically, Fig. \ref{fig:result_of_DeepPirates}(a) illustrates the estimation errors of the azimuth changes (in a period of 0.5s) predicted by a deep learning model when two persons move simultaneously in front of a PIR sensor. Fig. \ref{fig:result_of_DeepPirates}(b) illustrates the CDF of location errors in two testing scenarios whose sizes are $4m \times 4m$ and $7m \times 3.5m$ and with 4 PIR sensors deployed at the corners. The average localization errors in these two scenarios are 0.75m and 0.77m, respectively. It can be seen that, by leveraging this method, PIRNet tends to still work well in quite different deployment scenarios. In the future, we will conduct more experiments to further validate this method.
	
	\ifCLASSOPTIONcaptionsoff
	\newpage
	\fi

	\bibliographystyle{unsrt}
	\bibliography{PIRNet}

\end{document}